\documentclass[romanappendices]{IEEEtran} %[12pt] %used in draft 
\usepackage{amsmath,amssymb,verbatim,graphicx,rotating} 

\newtheorem{theorem}{Theorem}%[section]
\newtheorem{lemma}{Lemma}%[section]
\newtheorem{proposition}{Proposition}%[section]
\newtheorem{corollary}{Corollary}
\newtheorem{definition}{Definition}

\newcommand{\enproof}{\hfill $\Box$ \vspace*{1ex}}

\newcommand{\ssi}{} %{\mbox{}^{(i)}} %second submission
\newcommand{\mbssi}{} %{\mbox{}^{(i)}} %second submission

\newcommand{\mymathbb}[1]{{\mathbb{#1}}} 
\newcommand{\mymathsf}[1]{{\mathsf{#1}}} 

\newcommand{\dmn}{q} %{d}
\newcommand{\cA}{{\cal A}}

\newcommand{\cC}{{\cal C}}

\newcommand{\sD}{\mymathsf{D}}

\newcommand{\cG}{{\cal G}}

\newcommand{\ssf}{\mymathsf{f}}

\newcommand{\myF}{{\mymathbb{F}_{\dmn}}} %qkd
\newcommand{\myFnoarg}{{\mymathbb{F}}}
\newcommand{\myFk}{{\mymathbb{F}_{\dmn^k}}}

\newcommand{\cH}{{\cal H}}

\newcommand{\sS}{\mymathsf{S}}

\newcommand{\cL}{{\cal L}}

\newcommand{\cM}{{\cal M}}

\newcommand{\sQ}{\mymathsf{Q}}

\newcommand{\cV}{{\cal V}}

\newcommand{\cX}{{\cal X}}

\newcommand{\sW}{\mymathsf{W}}

\newcommand{\cZ}{{\cal Z}}

\newcommand{\vep}{\varepsilon}
\renewcommand{\phi}{\varphi} %qkd
\renewcommand{\subset}{\subseteq}
\renewcommand{\tilde}{\widetilde}
\renewcommand{\bar}{\overline}

\newcommand{\defeq}{\stackrel{\rm def}{=}}

\newcommand{\SNN}{\mymathbb{N}}

\newcommand{\Capa}{\mymathsf{C}} %{{\rm C}}
\newcommand{\transp}{\mbox{}^{\rm t}}

\newcommand{\tracenos}{{\rm Tr}\,}
\newcommand{\trace}{{\rm Tr}_{\myFk/\myF}\,}

\newcommand{\crd}[1]{|#1|}

\newcommand{\sypaltnoarg}{\mymathsf{f}_{\rm s}}
\newcommand{\sypalt}[2]{\sypaltnoarg(#1,  #2)}
\newcommand{\dpr}[2]{#1 \cdot #2}
\newcommand{\perpsypalt}{{\perp_{\rm s}}} %{{\perp_{\rm sp}}}

\newcommand{\spn}{\mymathsf{span}\,}
\newcommand{\Ccl}{C}
\newcommand{\Rqt}{R_{\rm q}} %{\kappa} %changed to \kcldual
\newcommand{\css}[2]{\mymathsf{S}_{\rm css}(#1,#2)}

\newcommand{\myFpower}[1]{\mymathbb{F}_{\dmn}^{#1}} %2nd_subm

\newcommand{\myFpowernoarg}[1]{\mymathbb{F}^{#1}}
\newcommand{\myFxpower}[1]{\myFpower{#1}}

\newcommand{\Qpl}[1]{\sQ^+} %{\sQ^+_{#1}}
\newcommand{\Qmi}[1]{\sQ^-}
\newcommand{\vart}{b} %{\rho} %{e} %{t} 

\newcommand{\Bp}[1]{G}
\newcommand{\zrv}{{\bf 0}}
\newcommand{\zrmat}{O} %
\newcommand{\Bsmall}{B} %{\cB} %change this to {B} later %used to be {\cB}
\newcommand{\Jgood}{\tilde{J}}

\newcommand{\crI}{J} %{\Gamma} %{J} 

\newcommand{\Jof}[1]{\prm(\Jgood)}

\newcommand{\prm}{\pi}

\newcommand{\chooses}[2]{\Big( \begin{array}{c} #1 \\[-1ex] #2 \end{array} \Big)}

\newcommand{\cst}[1]{\tilde{#1}} %{\hat{#1}}
\newcommand{\wght}{\sW}
\newcommand{\cwght}{\sW_{B}} %{\cst{\sW}}

\newcommand{\kperp}{\perp} %{\perp'}?
\newcommand{\embone}{\pi_1}
\newcommand{\embtwo}{\pi_2}
\newcommand{\Cone}{L_1}
\newcommand{\Ctwo}{L_2}
\newcommand{\tnum}{M}
\newcommand{\intint}[2]{\overline{[#1,#2]}} %{[#1,#2]_{\SINT}} %{\overline{[#1,#2]}} %{[#1,#2]'} %

\newcommand{\degF}{k} %{\nu}
\newcommand{\intk}{l} %{k}
\newcommand{\MatF}{\Phi}
\newcommand{\varKgen}{K}
\newcommand{\bsa}{\mymathsf{a}}
\newcommand{\bsb}{\mymathsf{b}}
\newcommand{\bab}[1]{\begin{matrix} | \\ #1 \\ | \end{matrix}}
\newcommand{\myFkpower}[1]{\myFnoarg_{\dmn^\degF}^{#1}}

\newcommand{\MatFa}{\MatF_{\bsa}}

\newcommand{\subgrp}{\le}

\newcommand{\CSone}{C_1}
\newcommand{\CStwo}{C_2^{\perp}} %in Shor's paper, $C_2$

\newcommand{\Noa}{N_{\rm o}}
\newcommand{\Koa}{K_{\rm o}}
\newcommand{\Roa}{R_{\rm o}}
\newcommand{\doa}{d_{\rm o}}
\newcommand{\doaj}{d_{{\rm o}}(j)}
\newcommand{\doaone}{d_{{\rm o}}(1)}
\newcommand{\doatwo}{d_{{\rm o}}(2)}

\newcommand{\HamW}{\mymathsf{w}}
\newcommand{\HamD}{\mymathsf{d}} %{D_{\rm H}}
\newcommand{\Dout}{d'} %{d_{\rm out}}
\newcommand{\Doutout}{d''} %{d_{\rm out}}
\newcommand{\inu}{\nu}
\newcommand{\dblith}{\mbox{}_{,\inu}}
\newcommand{\ith}{_{\inu}} %{(\inu)}
\newcommand{\dvsr}{A}

\newcommand{\gtN}{\gamma_k}

\newcommand{\gtNhat}{\hat{\gamma}}
\newcommand{\cdB}{B}
\newcommand{\KoD}{K}
\newcommand{\KoE}{K'}
\newcommand{\genC}{L} %{L}
\newcommand{\stn}{\sS_{\rm enl}} %{\hat{\sS}} %{\sS_{\rm sym}} %{\sE}
\newcommand{\stA}{M}
\newcommand{\matGpr}{W}

\newcommand{\myFkgen}{{\mymathbb{F}_{\dmn^k}}}

\newcommand{\zrvb}{{\bf 0}}
\newcommand{\Peone}{P_{{\rm e},1}}
\newcommand{\Petwo}{P_{{\rm e},2}}
\newcommand{\Pej}{P_{{\rm e},j}}

\newcommand{\Peinj}{P_j}
\newcommand{\Peonein}{P_{1}} %^{\rm in}} {\rm in},
\newcommand{\Petwoin}{P_{2}} %^{\rm in}} {\rm in},

\newcommand{\myg}{g^{1)}}
\newcommand{\mygpr}{g^{2)}}
\newcommand{\mybeta}{\beta^{1)}}
\newcommand{\mybetapr}{\beta^{2)}}
\newcommand{\mypsi}{\phi'}
\newcommand{\tlG}{M} %{\tilde{G}}

\title{Concatenated Quantum Codes Constructible in Polynomial Time:
Efficient Decoding and Error Correction}

\author{%
  Mitsuru Hamada, \ {\em Member, IEEE} %\affiref{TM}\affiref{PRESTO}
\thanks{%
The material in this paper was presented in part at the
IEEE Information Theory Workshop,
Chengdu, China, Oct.\ 2006.}
\thanks{The author is with 
Quantum Information Science Research Center,
               Tamagawa University Research Institute,
6-1-1 Tamagawa-gakuen, Machida, Tokyo 194-8610, Japan.
He was also with PRESTO, Japan Science and Technology Agency,
4-1-8 Honcho, Kawaguchi, Saitama, Japan.
E-mail: {\tt mitsuru@ieee.org}.}
}  

\begin{document}

\maketitle

\begin{abstract}
A method for concatenating quantum error-correcting codes
is presented.
The method is applicable to
a wide class of quantum error-correcting codes
known as Calderbank-Shor-Steane (CSS) codes.
As a result, codes that achieve a high rate in the Shannon theoretic
sense
and that are decodable in polynomial time are presented.
The rate is the highest among those known
to be achievable by CSS codes.
Moreover, the best known lower bound on the greatest minimum distance 
of codes constructible in polynomial time is improved
for a wide range.
\end{abstract}


\section{Introduction \label{ss:intro}}

In the past decades, great efforts have been made to extend
information theory and its ramifications to quantum theoretical
settings. 
In particular, quantum error correction has been
an attractive field for both physicists and coding theorists.
The most important class of quantum error-correcting codes (quantum codes)
would be that of symplectic codes (stabilizer codes)~\cite{crss97,crss98,gottesman96}. These codes have direct relations with codes over finite fields satisfying some simple constraints on orthogonality.
% with repect to a symplectic bilinear form. 
This has allowed us to utilize many results from coding theory.
For example, quantum codes constructible in polynomial time are presented in \cite{AshikhminLT01} based on developments of algebraic geometry codes.
In the present paper, 
we propose a method for concatenating quantum codes,
which will be obtained
by developing Forney's idea of concatenated codes~\cite{forney}. 
As applications, we will treat
two complexity issues on quantum codes to be described below.

The codes to be proposed in this paper fall in
the class called Calderbank-Shor-Steane 
(CSS) codes~\cite{CalderbankShor96,steane96a}
or a closely related code class.
CSS codes form a class of symplectic codes.
According to \cite[p.~2492, last paragraph]{steane99e},
a CSS quantum code is succinctly represented as
a pair of linear codes $(C_1,C_2)$
with $C_2^{\perp} \subgrp C_1$,
where $C^{\perp}$ denotes the dual of $C$,
and by $B \subgrp C$, we mean that
$B$ is a subgroup of an additive group $C$.
{\em In this paper,  any code pair 
written in the form $(C_1,C_2)$ is supposed to satisfy
the constraint $C_2^{\perp} \subgrp C_1$.}\/
Note that a CSS quantum code is a Hilbert space associated with
a code pair $(C_1,C_2)$ in the manner described in \cite{CalderbankShor96} with $\cC_1=C_1$ and $\cC_2=C_2^{\perp}$.
However, 
we will keep the style~\cite{steane99e} of not mentioning Hilbert spaces
as far as it is possible.
For the original purpose of quantum error correction, 
$C_1$ is used for bit-flip errors and $C_2$ for phase-shift errors.
Therefore, if codes $C_1$ and $C_2$ are both good,
the CSS quantum code specified by $C_1$ and $C_2$ is good.

This paper presents a method for 
creating code pairs, $(L_1,L_2)$, of relatively large lengths
by concatenating shorter code pairs.
%$(C_1\ssi,C_2\ssi)$ and $(D_1,D_2)$.  
%This will be acomplished by developing Forney's idea of concatenated codes~\cite{forney}. 
The main technical problem to be resolved in this work
is to concatenate code pairs
in such a way that the resulting pair $(L_1,L_2)$ satisfies
$L_2^{\perp} \subgrp L_1$. Our method for concatenation is applicable
to any combination of a $q$-ary inner code pair, $(C_1,C_2)$, and a $Q$-ary outer code pair, $(D_1,D_2)$,
as far as $q^k=Q$, where $k$ is the number of information digits
of the inner code pair $(C_1,C_2)$.
%, equals the size of the code alphabet of the outer code pair $(D_1,D_2)$.
This generality is the same as Forney's method has.

Using this general method, we give solutions
to two complexity issues on symplectic codes.
One issue is on decoding complexity, and the other on
complexity of code construction.
The ability of error correction will be measured in terms of
(i) the decoding error probability (as usual in Shannon theory)
for the first issue, 
%i.e., in treating the decoding complexity,
and in terms of (ii) the minimum distance (as usual in coding theory)
for the second.
Another related issue of construction complexity with (i) %criterion 
will be discussed elsewhere~\cite{hamada07expl}.

Regarding history of results on (i),
the existence of good 
CSS codes has been proved without regard to complexity issues.
%decoding or consruction complexity.
Specifically, 
the rate $1-2h(p)$,
where $h$ denotes the binary entropy function,
was called the Shannon rate in \cite{ShorPreskill00}
and a proof of the achievability of $1-2h(p)$ was given in \cite{hamada03s}, while the achievability of a smaller rate $1-2h(2p)$, $0 \le p < 1/2$, had been known~\cite{CalderbankShor96}.
Here, the channel
is $\mbox{BSC}(p)$,
the binary symmetric channel of the probability of flipping bits $p$.%
\footnote{The asymptotically good code pairs in \cite{CalderbankShor96,hamada03s} have form $(C,C)$. For a more detailed description of the previous result~\cite{hamada03s}, we need the definition of achievability in Section~\ref{ss:goal}.
The above description is for the simple case where $W_1=W_2=\mbox{BSC}(p)$
in the setting of Section~\ref{ss:goal}.%
}
If a wider class of quantum codes are considered,
higher rates are known to
be achievable (e.g., by symplectic codes~\cite{hamada02c} 
or Shannon-theoretic random codes~\cite{devetak03}).
However, 
%efficient decoding schemes for these codes are not known 
%for large code-lengths because 
none of these codes has a rich 
structure that allows efficient decoding.

In this paper, we consider the issue of constructing efficiently
decodable CSS codes.
%Our approach is that of concatenated codes~\cite{forney},
By the proposed method of concatenation,
we prove that the rate $1-2h(p)$ is achievable 
%with codes of polynomial decoding complexity.
with codes for which the error pattern can be estimated
%from the syndrome 
in polynomial time.

We remark another major approach, i.e., 
that of low-density parity-check or sparse-graph codes had already been
taken to construct CSS codes~\cite{MacKayMM04}.
However, they did not give asymptotically good
sparse-graph quantum codes but codes of particular lengths around $10^4$.
One of the authors~\cite{MacKayMM04} has even made a conjecture that
any dual-containing sparse-graph codes may be asymptotically bad;
note a dual-containing code $C$ corresponds to a pair $(C,C)$
in our notation.
Moreover,
the present work is different from \cite{MacKayMM04} in 
that the decoding error probability is evaluated without approximation
or resort to simulation.

In the latter half of the paper,
we will evaluate the minimum distance, (ii), of concatenated CSS codes
that are obtained with our general concatenation method.
The main result of this part (Theorem~\ref{prop:gen}) parallels 
a known lower bound~\cite{VladutKT84} to the largest minimum distance of
classical constructible codes to some extent.
%and also implies known lower bounds on CSS codes~\cite{ChenLX01}.

Regarding history of results on (ii),
the polynomial constructibility of classical codes was formulated 
and discussed in \cite{VladutKT84,TsfasmanVladut,stepanov}
with the criterion of minimum distance.
This problem formulation was brought into the realm of
quantum coding in \cite{AshikhminLT01}, 
which was followed by \cite{ChenLX01}.
We will evaluate the asymptotic relative minimum distance 
of concatenated CSS codes produced by the proposed method, 
%of concatenation, 
and compare these codes with known ones
%these codes with those in \cite{ChenLX01,AshikhminLT01}
%in \cite{hamada06ccc}
to show improvement for a wide range.
Furthermore, a code construction known as Steane's
enlargement of CSS codes is combined with the proposed concatenation method,
which will turn out to be effective.

The present work is motivated by the observation~\cite{ShorPreskill00}
(also described in \cite{hamada03s,hamada06s})
that good code pairs $(C_1,C_2)$, 
not the corresponding CSS quantum codes,
are useful
%in a converted form $C_1/C_2^{\perp}$, 
for quantum key distribution. We remark that 
for such cryptographic applications,
we need only 
classical information processing, not quantum information processing. 
For example, in a well-known application 
to quantum key distribution~\cite{ShorPreskill00},
we need quantum devices only for modulation.

Because of such background, the present work, in the previous version,
used
a formalism emphasizing cryptographic applications for presentation of results.
However, the author follows reviewers' comments
that the results should be presented in the context
of quantum error correction.
Still, the author remarks that the main result on efficient decoding
(Theorem~\ref{th:performance}) applies both to quantum error correction
and to communication over wiretapped channels.
Note that decoding (recovery operation) for a quantum code
is given as a completely positive linear map,
which is surely beyond classical information processing,
%, which is not classical information processing,
and even if one could find some non-CSS-type quantum codes
with efficient recovery operation,
it would not imply Theorem~\ref{th:performance}, which
claims that decoding 
of codes, $L_1/L_2^{\perp}$ 
and $L_2/L_1^{\perp}$, is
classical information processing of polynomial complexity.

The present paper was originally prepared as two seperate
manuscripts to treat the two issues respectively,
but they have been merged due to a request of the associate editor.
We remark that the part
treating the issue on minimum distance, starting
from Section~\ref{ss:mindis_intro},
can be read
independently from 
Sections~\ref{ss:parity_check_m} to \ref{ss:performance},
which treat the issue on decoding.

The remaining part of this paper is organized as follows.
In Section~\ref{ss:qc_cc}, we fix our notation. 
In Section~\ref{ss:goal}, a main statement on efficient decoding
is presented.
In Section~\ref{ss:ccc}, concatenated CSS codes are defined.
In Sections~\ref{ss:dec}--\ref{ss:synd_dec}, 
a method for decoding is described.
Specifically, a decoding strategy is described in Section~\ref{ss:dec},
a needed fundamental lemma is given and proved in Section~\ref{ss:ef} and \ref{app:proof_expansion}, respectively,
and syndrome decoding for concatenated CSS codes is described in detail in Section~\ref{ss:synd_dec}.
%***our construction is based.
The statement in Section~\ref{ss:goal} is proved in Section~\ref{ss:performance}.
In Section~\ref{ss:mindis_intro}, moving to the topic on (ii), 
a useful metric for quotient spaces is reviewed.
A basic lemma on the minimum distance of concatenated CSS codes is
presented in Section~\ref{ss:mindis},
and a general lower bound on the minimum distance is given
in Section~\ref{ss:bound}.
A restricted but more concrete bound is derived
from the general one in Section~\ref{ss:bound}
%in Sections~\ref{ss:clb} and \ref{ss:steane}
to show an improvement in Section~\ref{ss:cmp}.
In Section~\ref{ss:steane}, 
Steane's enlargement is combined with the concatenation method.
Section~\ref{ss:summary} contains a summary.

\section{Notation and Terminology \label{ss:qc_cc}}

%\subsection{Basic Notation}

%We fix some notation.
The set of consecutive integers 
$\{ l,l+1,\dots,m \}$ is denoted by $\intint{l}{m}$. 
%$=\{ l,l+1,\dots,m \}$
We use the dot product defined by
$ %begin{equation}\label{eq:dotpr}
\dpr{(x_1,\dots,x_n)}{(y_1,\dots,y_n)}=\sum_{i=1}^{n} x_iy_i
$ %end{equation}
on $\myFpowernoarg{n}$, where $\myFnoarg$ is a finite field.
For a subspace  %2ndsubm %subset 
$\Ccl$ of $\myFpowernoarg{n}$, 
$C^{\perp}$ denotes the usual dual $\{ y \in\myFpowernoarg{n} \mid \forall x\in C, \ \dpr{x}{y}=0 \}$. Similarly, 
$C^{\perpsypalt}$ denotes the dual 
$\{ y \in\myFpowernoarg{2n} \mid \forall x\in C, \ \sypalt{x}{y}=0 \}$
of $C$
with respect to the symplectic form $\sypaltnoarg$ defined below.
A subspace $\Ccl$ of $\myFpowernoarg{n}$ is called an $[n,k]$ code if
$k=\log_{\crd{\myFnoarg}} \crd{C}$.
As usual, $\lfloor a \rfloor$ denotes the largest
integer $a'$ with $a'\le a$, and $\lceil a \rceil = - \lfloor - a \rfloor$.
The transpose of a matrix $A$ is denoted by $A\transp$.
The juxtaposition of vectors $x_1, \dots, x_n$ from a linear space
is denoted by $(x_1|\cdots |x_n)$.
Throughout, 
we fix a finite field $\myF$ of $q$ elements, and
construct codes over $\myF$.

%\subsection{Steane's Notation}
In the sense of \cite{steane99e}, 
an $[[n,k]]$ symplectic quantum code (also known as a stabilizer code) 
can be viewed
as a subspace of $\myFxpower{2n}$ 
that contains its dual
with respect to the standard symplectic bilinear form $\sypaltnoarg$
defined by
\begin{equation*} %\label{eq:syp_def}
\sypaltnoarg\big( (u_x|u_z),(v_x|v_z)\big)
=\dpr{u_x}{v_z}-\dpr{u_z}{v_x}.
\end{equation*}
Such an $(n+k)$-dimensional subspace may be called
an $\sypaltnoarg$-dual-containing code,
but will be called an $[[n,k]]$ {\em symplectic code}\/ 
(over $\myF$) for simplicity in this paper.

We can also characterize symplectic codes with 
their generator matrices~\cite{steane99e}.
Namely, the subspace spanned by
the rows of a full-rank matrix of the form 
${\cal G} = [  G_x \, G_z ]$, 
where $G_x$ and $G_z$ are $(n+k)\times n$ matrices,
is a symplectic code if
$G_x$ and $G_z$ satisfy
\begin{equation*} %\label{eq:HG}
H_x G_z^{\rm t} - H_z G_x^{\rm t} = \zrmat
\end{equation*}
for some $(n-k)\times 2n$ full-rank matrix
$\cH=[H_x\, H_z]$ such that $\spn \cH \le \spn \cG$.
Here, $\zrmat$ denotes the zero matrix, and
$\spn A$ denotes the space spanned by the rows of $A$.
The space $\spn \cH$ is
the $\sypaltnoarg$-dual of $\spn \cG$.

We can say~\cite{steane99e} that 
the CSS code construction \cite{CalderbankShor96,steane96a} is to take
classical codes $C_1$ and $C_2$ with $C_1^{\perp} \subgrp C_2$,
and form
\begin{equation}
{\cal G} = \left[ \begin{array}{cc} G_1 & \zrmat \\ \zrmat & G_2 \end{array}
\right], \;\;\;\;\;
{\cal H} = \left[ \begin{array}{cc} H_2 & \zrmat \\ \zrmat & H_1 \end{array}
\right]
    \label{G1G2}
\end{equation}
where $G_i$ and $H_i$ are the classical generator and parity check matrices
of $C_i$.

We call
a pair of linear codes $(C_1,C_2)$, 
where $C_1,C_2\subgrp \myFpower{n}$,
satisfying the CSS constraint
\begin{equation}\label{eq:css_cond}
\CStwo \subgrp \CSone 
\end{equation}
and
\begin{equation} \label{eq:css_k}
k=\dim_{\myF} C_1 + \dim_{\myF} C_2 -n
\end{equation}
an $[[n,k]]$ {\em code pair}\/ over $\myF$.
The corresponding $[[n,k]]$ symplectic code is called
an $[[n,k]]$ CSS code and is denoted by $\css{C_1}{C_2}$.

The following slight generalization of linear codes is useful
for our argument. 
While we usually use a linear code, i.e., subspace of $\myFpower{n}$,
we also call an additive quotient group $C/B$ a code
($B\le C\le\myFpower{n}$).
If we need to distinguish codes of the form $C/B$
from ordinary linear codes, we will call 
$C/B$
a {\em quotient code}\/ over $\myF$.%
\iffalse
The (information) rate of the quotient code $C/B$ is defined as
$n^{-1}\log_{q} |C|/|B|$.%
\fi
\footnote{The quotient codes
can really be used for transmission of information in the following manner.
The sender encodes a message into a member $c$ of $C/B$, 
% a code-coset, so to speak, 
chooses a word in $c$
at random and then sends it through the channel.
%\iffalse
Clearly, if $C$ is $J$-correcting ($J\subset \myFpower{n}$) 
in the ordinary sense,
$C/B$ is $(J+B)$-correcting (since adding a word in $B$ to the `code-coset'
$c$ does not change it).
%\fi
This kind of schemes had been known to be useful for 
coding on wiretap channels~\cite{wyner75}.}

Using the structure of $C/B$ explicitly is especially
useful for describing correctable errors of quantum error-correcting
codes.
It is known that 
if the above code $\spn \cG$ is $\Gamma$-correcting 
%in the ordinary sense,
(i.e., if $y-x \notin \spn \cG$ for any $x,y \in \Gamma$ with $x \ne y$),
then the corresponding quantum error-correcting code is
$\cA$-correcting for $\cA$ consisting of the quantum error patterns
represented by the vectors in $\Gamma+\spn \cH$.
(This form of the basic fact
can be found in \cite[Lemma~2]{hamada02c}.)
Note that the set $\Gamma+\spn \cH \subset \myFpower{2n}$ is formally the same as
the correctable errors of the fictitious quotient code $\spn \cG/\spn \cH$,
which is also called a symplectic code.
%This reduces design issues of quantum codes to those of codes over finite fields. 
%in part (and this part is of our interest here). 

For the CSS construction, the set of correctable errors $\Gamma+\spn \cH$
can be written typically as follows.
If $C_i$ is $J_i$-correcting,
by (\ref{G1G2}), we can set
\begin{gather}
\Gamma+\spn \cH \notag\\
= \{ (x|z) \mid \mbox{$x \in J_1 + \spn H_2$
and
$z \in J_2 + \spn H_1$} \} \notag\\
= 
\{ (x|z) \mid \mbox{$x \in J_1 + C_2^{\perp}$ and
$z\in J_2 + C_1^{\perp}$} \}.  \label{eq:CSScorrectrableE}
\end{gather}
%Note that the CSS property (\ref{eq:C1C2}) is equivalent to $C_1^{\perp} \le C_2$.
The number $k/n$ is called the (information) rate of 
the code pair $(C_1,C_2)$, and equals
that of $C_1/C_2^{\perp}$ and that of $C_2/C_1^{\perp}$.

The condition (\ref{eq:css_cond}) is equivalent to
that $C_1^{\perp}$ and $C_2^{\perp}$ are orthogonal %perpendicular 
to each other.
Here, with two codes $C$ and $C'$ given, we say $C$ is orthogonal %perpendicular
to $C'$ 
and write
\[
C \perp C'
\]
if $\dpr{x}{y}=0$ for any $x \in C$ and $y\in C'$.
Note that $C \perp C'$ 
if and only if (iff) $C'\le C^{\perp}$, or equivalently, 
iff $C\le C'\mbox{}^{\perp}$.

\section{Theorem on Efficient Decoding \label{ss:goal}}

\subsection{Main Theorem on Efficient Decoding}

The first goal in this paper %, in a long span, 
is to find a code pair $(L_1,L_2)$
such that
both $L_1/L_2^{\perp}$ and $L_2/L_1^{\perp}$ 
have small decoding error probabilities
and are decodable with polynomial complexity.

In particular, we will explore the achievable rates
of efficiently decodable quotient codes.
Here, given a sequence of code pairs $\{ (L_{1,\nu}, L_{2,\nu}) = (L_{1},L_{2}) \}$ and a pair of memoryless additive channels $(W_1,W_2)$,
we say $\{ (L_{1},L_{2}) \}$ {\em achieves}\/ a rate $R$ for $(W_1,W_2)$
if the rate of $L_{1}/L_{2}^{\perp}$ approaches $R$
and the decoding error probability of 
$L_{1}/L_{2}^{\perp}$ and that of $L_{2}/L_{1}^{\perp}$ both go to zero;
a memoryless additive channel $W$ actually denotes 
the channel specified by a probability distribution $W$ on $\myF$; 
this channel changes an input $a \in\myF$ into $b$ with probability $W(b-a)$. 
The first half of this paper is devoted to proving the following theorem. 
%on the Shannon theoretic criterion.
\begin{theorem}\label{th:performance}
Assume we are given
a pair of memoryless additive channels $W_1,W_2$, and
we have a sequence of $[[n,k]]$ code pairs 
$(C_1,C_2)$ over $\myF$
whose decoding error probabilities, $\Peonein$ for $C_1/C_2^{\perp}$ 
and $\Petwoin$ for $C_2/C_1^{\perp}$, are bounded by
\begin{equation} \label{eq:bound_inner}
P_j \le q^{-n  E(W_j,r_j) +o(n)}, \quad n\in \SNN,\, j=1,2.
%\Pein=\max\{\Peonein, \Petwoin \}\le a_n  q^{-n \min\{ E(W_1,r_1)}, E(W_2,r_2) \} }.
\end{equation}
Here, $r_j$ is the rate of $C_j$ (when it is viewed as a classical code). 
Then, for any fixed number $\Roa$, $0 < \Roa \le 1$,
there exists a sequence of
$[[\Noa,\Koa]]$ code pairs $(L_1,L_2)$ of the following properties.
(i) The rate $\Koa/\Noa$ approaches $\Roa$. 
(ii) The decoding error probability $\Pej$
is bounded by 
\begin{gather*}\label{eq:exp_ccc}
\limsup_{\Noa\to\infty} - \frac{1}{\Noa} \log_q \Pej \\
\ge \frac{1}{2} \, \max_{(r_1+r_2-1)(R_1+R_2-1)=\Roa}\, 
\min_{j\in \{1,2\}} (1-R_j)E(W_j,r_j)
\end{gather*}
for $j=1,2$, where the maximum 
%in the expression of the bound
is taken over $\{ (r_1,r_2,R_1,R_2) \mid 0 \le r_j \le 1,  0 \le R_j \le 1 \mbox{ for $j=1,2$}$, $(r_1+r_2-1)(R_1+R_2-1)=\Roa \}$. %4LAYOUT
(iii)  The codes $L_1/L_2^{\perp}$ and $L_2/L_1^{\perp}$ are decodable
with algorithms of polynomial complexity.
\end{theorem}

In the theorem, the sequence $\{ (L_1,L_2) \}$ actually
consists of $[[N_{{\rm o},\nu},K_{{\rm o},\nu}]]$
code pairs $(L_{1,\nu},L_{2,\nu})$, $\nu\in\SNN$, such that $N_{{\rm o},\nu} \to \infty$
as $\nu$. % \to \infty$.

To prove this theorem, %constructing concatenated code pairs
we will present a general concatenation method
for CSS codes.
Then, proving (i) and (ii) will be a routine, following \cite{forney}. 
However, to establish (iii),
%besides the method for concatenation, 
a method for constructing parity check matrices that enables us
to decode $L_j/L_{\bar{j}}$, where $\bar{1}=2$ and $\bar{2}=1$,
in polynomial time is needed.
This will also be presented, %in what follows. %,
and besides 
the concatenation method, this would be the most novel part of the present work.

\subsection{Review of Needed Results on Exponential Error Bounds}

To make Theorem~\ref{th:performance} meaningful, 
we need good codes satisfying the premise of the theorem.
These codes will be used as inner codes in concatenation.
Therefore, we begin with reviewing results on the needed good inner codes~\cite{hamada05qc}. 

We know the existence of a sequence of $[[n,k]]$ code pairs $(C_1,C_2)$
attaining the random coding error exponent $E_{\rm r}(W_j,r_j)$: 
For any rate pair $(r_1,r_2)$ and
for any pair of additive channels $(W_1,W_2)$, we have
\begin{equation*} 
P_j \le q^{-n  E_{\rm r}(W_j,r_j) +o(n)}, \quad n\in \SNN,\, j=1,2
%\Pein=\max\{\Peonein, \Petwoin \}\le a_n  q^{-n \min\{ E(W_1,r_1)}, E(W_2,r_2) \} }.
\end{equation*}
% as in (\ref{eq:bound_inner}).
where
\begin{equation}\label{eq:Er}
E_{\rm r}(W_j, r_j) = \min_{Q} [D(Q||W_j) + |1- r_j -H(Q)|^+].
\end{equation}
%for additive memoryless channel $W_j$ of alphabet $\myF$,
Here, $H$ and $D$ denote
the Shannon entropy 
and the Kullback-Leibler information, respectively,
the minimum is taken over all probability distributions 
on $\myF$, and $|x|^+=\max\{ 0, x\}$.

This was proved as follows~\cite[Section~10.3]{hamada05qc}. We know there exists a good classical code $C_1$ 
satisfying (\ref{eq:bound_inner}) 
for $j=1$ with $E=E_{\rm r}$. 
Then, for an arbitrarily fixed $n$,
we consider all possible codes $C_2$ with $C_1^{\perp} \subgrp C_2$
of a fixed size. Evaluating the average of decoding error probability
of $C_2/C_1^{\perp}$ over this ensemble, 
we obtain (\ref{eq:bound_inner}) also for $j=2$.

\subsection{Achievable Rates of Efficiently Decodable CSS Codes}
%Concatenated Codes of CSS Type 
%\label{ss:ach}}
%and Decoding Complexity 

We describe implications of Theorem~\ref{th:performance} here. 
As reviewed above,
%in Section~\ref{ss:goal}, 
the bound in (\ref{eq:bound_inner}) has been proved for 
the random coding exponent $E=E_{\rm r}$.
Note in this case,
$E(r_j,W_j)$ is positive whenever
$r_j < \Capa(W_j)=1-H(W_j)$, $j=1,2$, and that
for any $\vep$, we can take $r_1,r_2,R_1,R_2$ 
such that $\Capa(W_j)>r_j > \Capa(W_j) -\vep$ and $1 > R_j >1-\vep$ for $j=1,2$.
Hence, for any $\delta>0$, 
we can choose $r_1,r_2,R_1,R_2$
such that $\Roa=(r_1+r_2-1)(R_1+R_2-1) > \Capa(W_1)-\Capa(W_2) -1 -\delta$ 
and $\min_{j\in \{1,2\}} (1-R_j)E(W_j,r_j)$ is positive.
Thus, the rate $\Capa(W_1)+\Capa(W_2)-1$ is achievable.
In the literature, e.g., in \cite{MacKayMM04},
the binary case ($q=2$) with $W_1=W_2$
has sometimes been discussed {\em without}\/ presenting efficiently decodable codes that achieve any positive rate.
In this binary case, some call
$1-2H(W_1)$ the Shannon rate, which equals the rate 
$\Capa(W_1)+\Capa(W_2)-1= 1- H(W_1) -H(W_2)$ for $W_1=W_2$.
This rate is the highest among those known
to be achievable by CSS codes.

The pair of efficient decoders for $L_1/L_2^{\perp}$ and $L_2/L_1^{\perp}$ (Theorem~\ref{th:performance}), 
which involve only with classical information processing, 
will be useful for quantum error correction
provided the recovery operation is done
in a standard manner~\cite{CalderbankShor96,steane96a},
i.e., by measuring the syndromes and applying 
the inverse of the estimated quantum error pattern. The task of 
the above classical decoders is estimating the error pattern from the syndromes.

We remark that Theorem~\ref{th:performance}
%our evaluations on the reliability of code pairs
%$(L_1,L_2)$ 
has direct implications on the reliability %(and security)
of the CSS quantum codes specified
%as in the footnote in Section~\ref{ss:intro},
by $(L_1,L_2)$:
%, which are involved with quantum mechanical operations:
The fidelity of the CSS code is lower-bounded by $1-\Peone-\Petwo$
owing to (\ref{eq:CSScorrectrableE}).
%(see, e.g., \cite{hamada06s,hamada05qc}).
%(Section~\ref{ss:qc_cc}).

\iffalse
The literature has been lacking
%cryptographic (quotient) codes 
codes of the CSS type
that allow efficient decoding and
achieve the rate $1-2h(p)$, %~\cite{hamada03s},
% or smaller $1-2h(2p)$~\cite{CalderbankShor96},
which is the goal of this paper.
\fi

\section{Concatenation of Codes of CSS Type \label{ss:ccc}}

\subsection{Construction of Codes}

In this section,
we will present a method  
for creating concatenated code pairs, $(L_1,L_2)$
with $L_1^{\perp} \subgrp L_2$.

\begin{lemma}\label{prop:gg'}
Assume $(C_1, C_2)$ is an $[[n,k]]$ code pair over $\myF$,
and
\begin{equation*}
C_1=C_2^{\perp}+\spn\{ \myg_1,\dots,\myg_k \}.
%%%C_2^{\perp}&=&\spn\{ h_1,h_2,\dots,h_{n-k_2}\}\\
\end{equation*}
Then, we can find vectors $\mygpr_1,\dots,\mygpr_k$ such that
\begin{equation*}
%C_1^{\perp}&=&\spn\{ h'_1,h'_2,\dots,h'_{n-k_1}\}\\
C_2=C_1^{\perp}+\spn\{ \mygpr_1,\dots,\mygpr_k \}
\end{equation*}
and
\begin{equation}\label{eq:gg'}
\dpr{\myg_i}{\mygpr_j} = \delta_{ij}
\end{equation}
where $\delta_{ij}$ is the Kronecker delta.
\end{lemma}

\begin{figure} %[h]
\begin{center}
\includegraphics[scale=0.8]{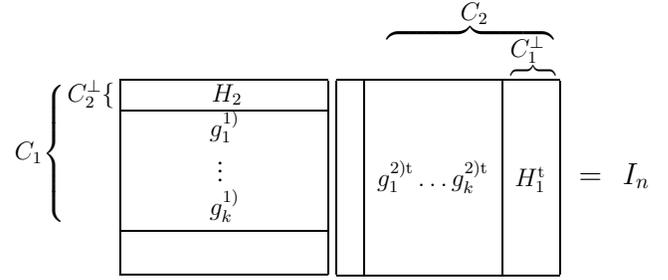} % Here is how to import EPS art
%\vspace*{-2ex}
\caption{A basic structure of an $[[n,k]]$ code pair.
\label{fig:css}}
\end{center}
\end{figure}

{\em Proof.}\/ 
See Fig.~\ref{fig:css}.
If $C_1=C_2^{\perp}+\spn\{ \myg_1,\dots,\myg_k \} \le \myFpower{n}$ 
and $H_2$ is a full-rank parity check matrix
of $C_2$, we have an invertible matrix, $A$, as depicted at the left-most
position of Fig.~\ref{fig:css}.
Of course, we have its inverse $A^{-1}$, which is depicted next to $A$
in the figure.
%Write $\mygpr_1\transp,\ldots,\mygpr_k\transp$ for the $(n-k_2+1)$-th to 
Write $g^{2){\rm t}}_1, \ldots, g^{2){\rm t}}_k$ for the $(n-k_2+1)$-th to 
$k_1$-th columns of $A^{-1}$.
Then, we see that $\dpr{\myg_i}{\mygpr_j} = \delta_{ij}$ and the last $n-k_1$ columns 
of the second matrix are
orthogonal %perpendicular 
to the $[n,k_1]$ code $C_1$. 
\mbox{} \enproof %LAYOUT

Let $(C_1,C_2)$ be an $[[n,k]]$ code pair over $\myF$,
where $C_1$ and $C_2$ are an $[n,k_1]$ code and an $[n,k_2]$ code, respectively,
with
$
k=k_1+k_2-n$.
Assume $\myg_i$ and $\mygpr_j$ satisfy
the conditions in Lemma~\ref{prop:gg'}. 
%Let $(D_1,D_2)$ be an $[[N,K]]$ code pair over $\myFk$.
The field $\myFk$ is an $\myF$-linear vector space, 
and we can take bases $\big( \mybeta_j \big)_{j=1}^{k}$ and
$\big( \mybetapr_j \big)_{j=1}^{k}$ 
that are dual to each other 
with respect to the $\myF$-bilinear form (Section~\ref{app:proof_expansion} or, e.g., 
\cite{LidlNied,sloane})
%,stichtenoth}) %4LAYOUT
defined by
\begin{equation}\label{eq:f}
\begin{array}{ccl}
\ssf_{\rm t} &:& \myFk\times\myFk \to \myF,\\
&& (x,y) \mapsto \trace xy.
\end{array}
\end{equation}
Namely, we have bases $\big(\mybeta_j \big)_{j=1}^{k}$ and $\big( \mybetapr_j \big)_{j=1}^{k}$ 
that satisfy
\[
\ssf_{\rm t}\big(\mybeta_i,\mybetapr_j \big)
=\trace \mybeta_i\mybetapr_j 
=\delta_{ij}.
\]

Relating $\big(\myg_i,\mygpr_j\big)$ with $\big(\mybeta_i,\mybetapr_j\big)$ naturally, we 
have a map that sends vectors in $\myFkpower{N}$
to the space 
\[
\bigoplus_{l=1}^N \spn\{ g^{m)}_1,\dots,g^{m)}_k \}
\]
and that preserves the inner product.
Namely, applying 
\begin{equation}\label{eq:pi_m}
\begin{array}{ccc}
\pi_m &:& \myFk \to \spn \{ g^{m)}_1,\dots,g^{m)}_k \} \simeq C_{m}/C_{\bar{m}}^{\perp},\\
&& \sum_j z_j \beta^{m)}_j \mapsto \sum_j z_j g^{m)}_j
\end{array}
\end{equation}
to each coordinate of a vector 
\[
x=(x_1,\dots,x_N) \in \myFkpower{N},
\]
we have a vector in $\myFpower{nN}$ ($m=1,2$). This extension of $\pi_m$ is
again denoted by $\pi_m$:
\begin{eqnarray*}
\pi_m(x) = \big(\pi_m(x_1)|\cdots|\pi_m(x_N) \big).
\end{eqnarray*}
%where $(y_1|\cdots|y_N)$ denotes the juxtaposition.

Then, for any $x=(x_1,\dots,x_N)$ and 
$y=(y_1,\dots,y_N)$, 
%we have
\begin{equation}
\trace x \cdot y = \dpr{\embone(x)}{\embtwo(y)}. \label{eq:preserveip}
\end{equation}
This is because we have
\[
\trace x_i y_i = \dpr{\embone(x_i)}{\embtwo(y_i)}
\]
for each $i \in \intint{1}{N}$.

\begin{definition}\label{def:1}
The concatenation (or concatenated code pair made) 
of the generic $[[n,k]]$ code pair $(C_1,C_2)$ over $\myF$
and 
an $[[N,K]]$ code pair $(D_1,D_2)$ over $\myFk$
is the $[[nN, kK]]$ code pair
\[
(\embone(D_1)+\overline{C_2^{\perp}}, [\embone(D_2^{\kperp})+\overline{C_2^{\perp}}]^{\perp})
\]
over $\myF$,
where 
%$D_2^{\kperp}$ is the dual of $D_2$ with respect to the standard 
%inner product over $\myFk$, and 
\[
\overline{C_m^{\perp}}=\bigoplus_{i=1}^N C_m^{\perp}, \quad m=1,2.
\]
\end{definition}

The codes $C_1,C_2$ are sometimes called inner codes,
and $D_1,D_2$ outer codes.

\begin{theorem}\label{th:duals_ccc}
\[
[\embone(D_2^{\kperp})+\overline{C_2^{\perp}}]^{\perp} = \embtwo(D_2)+\overline{C_1^{\perp}},
\]
\[
[\embtwo(D_1^{\kperp})+\overline{C_1^{\perp}}]^{\perp} = \embone(D_1)+\overline{C_2^{\perp}}.
\]
\mbox{}
\end{theorem}

\begin{corollary}
The concatenated code pair in Definition~\ref{def:1} 
can be written as
\[
(\embone(D_1)+\overline{C_2^{\perp}}, \embtwo(D_2)+\overline{C_1^{\perp}}).
\]
\mbox{}
\end{corollary}

{\em Proof.}\/ It is enough to prove the second equality by virtue of 
the symmetry. 
First, we show
\begin{equation}\label{eq:2prove}
[\embtwo(D_1^{\kperp})+\overline{C_1^{\perp}}]^{\perp} \ge \embone(D_1)+\overline{C_2^{\perp}},
\end{equation}
which is equivalent to 
\[
\embone(D_1)+\overline{C_2^{\perp}} \perp
\embtwo(D_1^{\kperp})+\overline{C_1^{\perp}}.
\]
The code $\embone(D_1)$ is orthogonal %perpendicular 
to $\embtwo(D_1^{\kperp})$
by (\ref{eq:preserveip}), 
and to $\overline{C_1^{\perp}}$ trivially.
Similarly, $\overline{C_2^{\perp}}$ is orthogonal %perpendicular
to $\embtwo(D_1^{\kperp})$.
By the basic property (\ref{eq:css_cond}), $C_2^{\perp}$ and
$C_1^{\perp}$ are orthogonal %perpendicular 
to each other,
and hence, $\overline{C_2^{\perp}}$ is orthogonal %perpendicular 
to
$\overline{C_1^{\perp}}$.

Thus, we have (\ref{eq:2prove}).
Since 
$\dim_{\myF}[\embtwo(D_1^{\kperp})+\overline{C_1^{\perp}}]+ \dim_{\myF} [\embone(D_1)+\overline{C_2^{\perp}}]=nN$,
we have the lemma,  and hence, the corollary.
\enproof

%\begin{corollary}
%\end{corollary}

\subsection{Parity Check Matrices \label{ss:parity_check_m}}

Note that a generator matrix of 
$\embtwo(D_1^{\perp})+\overline{C_1^{\perp}}$ over $\myF$
has the form 
\begin{equation}\label{eq:H4cat}
H_{\rm o}=
\begin{bmatrix}
H_1 &O & \dots & O \\
O & H_1 & & O \\
\vdots &  &  \begin{rotate}{45}$\vdots$\end{rotate}  &   \\
%\vdots &  &  !LATER \vdots ROTATE! &   \\
O & O & & H_1 \\
G'_{1,1} & G'_{1,2} & \cdots & G'_{1,N} \\
\vdots &  \vdots  &        & \vdots \\
G'_{\tnum,1} & G'_{\tnum,2} & \cdots & G'_{\tnum,N}
\end{bmatrix}
\end{equation}
where $H_1$ is a parity check matrix of $C_1$, 
$O$ is the zero matrix,
$\tnum=N-K_1$ ($K_1$ is the dimension of $D_1$), 
and for each $(i,j)$, $G'_{j,i}$ is
a $k \times n$ matrix whose rows are spanned by $\mygpr_l$.
Hence, by Theorem~\ref{th:duals_ccc}, (\ref{eq:H4cat}) is a parity check matrix
of $\embone(D_1)+\overline{C_2^{\perp}}$. 

The next task is to devise a method to choose $G'_{j,i}$ 
in such a way that efficient decoding is possible.
We will present such a method below.

In the method, the matrices
$G'_{j,i}$ in (\ref{eq:H4cat}) are obtained from a parity check matrix 
$H=[h_{ji}]$ of $D_1$.
Recall we have fixed two bases 
$\bsb=\big( \mybeta_j \big)_{j=1}^{k}$ and
$\bsb'= \big( \mybetapr_j \big)_{j=1}^{k}$ that are dual to each other
in constructing concatenated codes. 
Take a root $\alpha$ of a primitive polynomial $f$ over $\myF$.
We set $\Phi(\alpha^i)=T^i$ for $i=0,\dots,q^k-2$,
where $T$ is the companion matrix of $f$, which will
be defined in Section~\ref{app:proof_expansion}, and put
$\Phi(0)=O$.
%, the zero matrix.
For simplicity, we set $\bsb=(1,\alpha,\dots,\alpha^{k-1})$.
(This basis will appear as $\bsa=(1,\alpha,\dots,\alpha^{k-1})$ in what follows.)

{\em Procedure for creating $G'_{j,i}$,
$j\in\intint{1}{M}, i\in\intint{1}{N}$}.
\begin{quotation}
\noindent
{\bf Step 1}. We produce $\Phi(h_{ji})$ from $h_{ji}$.

\noindent
{\bf Step 2}. We replace each row $\eta=(\eta_1,\dots,\eta_{k})$ of
$\MatF(h_{ji})$ by
\begin{equation}\label{eq:catG}
\sum_{m=1}^{k} \eta_m \mygpr_m\mbssi ,
\end{equation}
and set the resulting $k \times n$ matrix  
equal to $G'_{j,i}$.
\end{quotation}

{\em Example 1.}\/
(a) Let $q=2$ and $k=3$. The companion matrix of
%$f(\alpha)=\alpha^3+\alpha+1$
a primitive polynomial $f(x)=x^3+x+1$ is
\begin{equation*} 
T=\begin{bmatrix} 
0 & 0 & 1 \\
1 & 0 & 1 \\
0 & 1 & 0
\end{bmatrix}.
\end{equation*}
Let $\alpha$ be a root of $f(x)$, and $H=[ 1 \ \alpha]$
a parity check matrix of a code $D_1$ over $\myFk$.
Then, we have
\begin{equation*} 
H'= [\Phi(1) \ \Phi(\alpha) ] = \begin{bmatrix} 
1& 0 & 0 & 0 & 0 & 1 \\
0& 1 & 0 & 1 & 0 & 1 \\
0& 0 & 1 & 0 & 1 & 0
\end{bmatrix}.
\end{equation*}

(b) The parity check matrix $H_{\rm o}$ of $L_1$ in (\ref{eq:H4cat})
for the concatenation $(L_1,L_2)$ of an arbitrary $(C_1,C_2)$ and, say,
$(D_1,\myFkpower{2})$ should be obtained from $H'= [\Phi(1) \ \Phi(\alpha) ]$ by the additional process of Step 2 for our purpose.
While there are many parity check matrices of $L_1$ such as obtained
by row permutations from this matrix $H_{\rm o}$,
this particular choice of $H_{\rm o}$ gives the desired parity check matrix
of $L_1$, which is useful for efficient decoding.
\enproof

We will see how this method works
in Sections~\ref{ss:dec} through \ref{ss:synd_dec}.

\section{Decoding Strategy for Concatenated Codes of CSS Type\label{ss:dec}}

We first sketch how to decode the concatenated code
$\Cone/\Ctwo^{\perp}$,
where
$\Cone=\embone(D_1)+\overline{C_2^{\perp}}$ 
and $\Ctwo=[\embone(D_2^{\kperp})+\overline{C_2^{\perp}}]^{\perp} =\embtwo(D_2)+\overline{C_1^{\perp}}$.
This is a half of the pair 
$(\Cone/\Ctwo^{\perp},\Ctwo/\Cone^{\perp})$,
and the other half, having the same form, can be treated similarly.

%\footnote{%
We remark that in known applications of code pairs $(C_1,C_2)$ with $C_2^{\perp} \subgrp C_1$, i.e.,
for CSS quantum codes and cryptographic codes as in 
\cite{ShorPreskill00,hamada03s}, %,hamada06s},
the decoding should be a {\em syndrome decoding}, which consists of
measuring the syndrome, estimating the error pattern, and 
canceling the effect of the error.

We decode the code in the following two stages.
\begin{enumerate}
\item For each of the inner codes, $C_1/C_2^{\perp}$, we perform
a syndrome decoding.
%(as described in Sections~2 and 3 of \cite{hamada05qc} for preciseness).
\item For the outer code $D_1$, we perform an efficient decoding such as
bounded distance decoding. 
\end{enumerate}

For efficient decoding, the outer code
$D_1$ should allow a decoding algorithm of polynomial complexity in $N$.
%Assume $n\ssi=n$ for all $i$ for simplicity.
Then, if $N \ge q^{\tau k}$ and $k/n \to r$ as $n\to \infty$,
where $\tau>0$ and $r \ge 0$ are constants,
the concatenated codes $\Cone/\Ctwo^{\perp}$ 
can be decoded with polynomial complexity in $N$,
and hence in the overall code-length $nN$.
%(Note that the length $n$ of inner codes has order $O(N)$.) 
Generalized Reed-Solomon (GRS) codes~\cite{sloane} are examples of such codes.
%An $[N,K_1]$ GRS code have minimum distance $N-K_1+1$.
%LATER more examples, elsewhere

\begin{comment}
Now assume the sender sent a word $x \in (\myFpower{n})^N$,
$x$ suffered an additive error $e=(e_1,\dots,e_N)\in (\myFpower{n})^N$,
and the receiver received a word $y=x+e\in (\myFpower{n})^N$.
Using the upper half of 
the parity check matrix in (\ref{eq:H4cat}), where $H_1$ is involved,
the receiver decodes the inner quotient codes.
Namely, receiver estimates $e_i$, 
and subtract $\hat{e}=(\hat{e}_1,\ldots, \hat{e}_N)$ from $y$,
where $\hat{e}_i$ is the estimate of $e_i$, which is
a function of the measured syndrome.
The decoding error for $C_1/C_2^{\perp}$ occurs only if
$e_i$ is outside $\tilde{\crI}=J+C_2^{\perp}$,
where $C_1$ is $J$-correcting. 
At this stage, the received word $y$ can be changed into
the interim estimate
\[
y'=y-\hat{e}=x+(e-\hat{e}).
\]

We employ bounded distance decoding here for simplicity,
though other schemes for classical concatenated codes, 
such as generalized minimum distance (GMD) decoding~\cite{forney}, 
are also applicable.
Then, the error $e$ is correctable 
if $e$ is such that the number of inner codes with erroneous decoding 
(the number of $i$ with $e_i \ne \hat{e}_i$)
is less than $\vart$, 
where we assume
the outer code $D_1$ is $\vart$-error-correcting.
% with an efficient algorithm.
\end{comment}

The decoding for the outer code should be done based on
the latter half of the syndrome that comes from
the lower half of the parity check matrix in (\ref{eq:H4cat}).
This is possible as will be argued in Section~\ref{ss:synd_dec_cccss}.
For this argument, we need some lemma,
which is given in Section~\ref{ss:ef}. % and \ref{ss:synd_n_k}.

%\section{Basics on Codes over Extension Fields and Dual Bases \label{ss:ef}}
\section{Dual Bases and Homomorphisms of Extension Field Into Space of Matrices  \label{ss:ef}}

%The finite field $\myFk$ is a $\degF$-dimensional vector space over $\myF$.
If $\bsb=(\beta_j)_{j=1}^{k}$ is a basis of the $\myF$-linear vector
space $\myFk$, any element $\xi\in\myFk$ can be written as
\[
\xi= x_1 \beta_1 + \cdots + x_k \beta_k.
\]
The row
vector $(x_1,\dots,x_{\degF})$ obtained in this way is 
denoted by $\varphi_{\bsb}(\xi)$.
The next lemma is fundamental to
our arguments in what follows.
\begin{lemma}\label{lem:expansion}
Let $\bsa$ denote the basis
$(\alpha^{j-1})_{j=1}^{k}$ for a primitive element
$\alpha$ of $\myFk$,
and $\bsa'$ the dual basis of $\bsa$.
There exists a one-to-one map
$\MatFa:\myFk\to\myFpower{k\times k}$ (the set of $k\times k$ matrices over
$\myF$)
with the following properties.
For any $\xi,\xi'\in\myFkgen$,
\begin{equation}\label{eq:Tbasic}
\MatFa(\xi) \varphi_{\bsa}(\xi')\transp  = \varphi_{\bsa}(\xi\xi')\transp,\quad
\varphi_{\bsa'}(\xi) \MatFa(\xi')  = \varphi_{\bsa'}(\xi\xi')
\end{equation}
and 
\begin{equation}\label{eq:Tbasic2}
\MatFa(\xi)\MatFa(\xi')=\MatFa(\xi\xi'),\,\,\,
\MatFa(\xi)+\MatFa(\xi')=\MatFa(\xi+\xi').
\end{equation}
\end{lemma}

The lemma %and its remark are 
is proved in an elementary manner 
in Section~\ref{app:proof_expansion}.
The part of Lemma~\ref{lem:expansion}
only involved with $\varphi_{\bsa}$
has sometimes been used in implementing codes.
%implementing codes over extension fields~(e.g., \cite[Chap.~5]{ash}.) 
However, Lemma~\ref{lem:expansion}, 
in which dual bases $\varphi_{\bsa}$ and $\varphi_{\bsa'}$ are featured,
was devised here for decoding of concatenated code pairs.

\section{Proof of Lemma~\protect\ref{lem:expansion} \label{app:proof_expansion}}
%In this section,
We will first construct maps $\varphi_{\bsa}$ and $\MatFa$ satisfying (\ref{eq:Tbasic}) 
and (\ref{eq:Tbasic2}) 
except
`$\varphi_{\bsa'}(\xi) \MatFa(\xi')  = \varphi_{\bsa'}(\xi\xi')$', 
%in Section~\ref{ss:ff},
and move on to proving the remaining part of the lemma.
% in Section~\ref{ss:dualbasis}.

\subsection{Companion Matrix \label{ss:ff}}

We use the following alternative visual notation for $\varphi_{\bsb}$ 
in the case where $\bsb=\bsa$:
\[
\begin{matrix}
|\\ \xi\\| 
\end{matrix}
= \varphi_{\bsa}(\xi)\transp \quad \mbox{which has form} \quad
\begin{bmatrix}
\xi_0\\ \vdots \\ \xi_{\degF-1}
\end{bmatrix}.
\]

Let $f(x)=x^{\degF}-f_{\degF-1}x^{\degF-1}- \cdots-f_1 x -f_0$
be the minimum polynomial of $\alpha$ over $\myF$.
%Then, $g(x)$ is  a primitive irreducible polynomial over $\myF$.
The companion matrix of $f(x)$ is 
\begin{equation}\label{eq:cm}
T=\begin{bmatrix} 
&0_{\degF-1} & f_0 \\
&{\Large I_{\degF-1}} & \begin{matrix} f_1 \\ \vdots \\ f_{\degF-1} \end{matrix}
\end{bmatrix}
\end{equation}
where $0_{\degF-1}$ is the zero vector in $\myFpower{\degF-1}$,
and $I_{\degF-1}$ is the $(\degF-1) \times (\degF-1)$ identity matrix.
Note that
\begin{equation}\label{eq:Ta}
T 
= \begin{bmatrix} | & & | \\
                     \alpha^{1} & \cdots & \alpha^{\degF}\\
                      | & & | 
       \end{bmatrix}.  %,\quad i,j\in\intint{0}{\dmn^\degF -2}
\end{equation}
Then, 
we have
\begin{equation}\label{eq:Tb}
T \begin{matrix}
|\\ \alpha^i \\| 
\end{matrix}
= \begin{matrix} | \\
                     \alpha^{i+1}\\
                  |
       \end{matrix},
\quad i\in\intint{0}{\dmn^\degF -2},
\end{equation}
%for any $\s$i\in\intint{0}{\dmn^\degF -2}$. 
which can easily be checked.

\begin{comment}
{\em Proof of (\ref{eq:Tb}).}\/
%Let $\xi \myFk$ and $\varphi_{\bsa}(\xi)=(x_1,\dots,x_\degF)$.
Let $\varphi_{\bsa}(\alpha^i)=(x_1,\dots,x_\degF)$.
Then,
\[
T \varphi_{\bsa}(\alpha^i)\transp = \sum_{j=1}^{\degF} x_j \bab{\alpha^{j}} 
\]
by (\ref{eq:Ta}). The right-hand side can be written as
$\sum_{j=1}^n x_j \varphi_{\bsa}(\alpha^j)\transp=
\varphi_{\bsa}(\sum_{j=1}^n x_j \alpha^j)\transp=
\varphi_{\bsa}(\alpha\sum_{j=1}^n x_j \alpha^{j-1})\transp=
\varphi_{\bsa}(\alpha\alpha^{i})\transp$, completing the proof.
\enproof
\end{comment}

We list properties of $T$, all of which easily 
follow from (\ref{eq:Tb}). 
By repeated use of (\ref{eq:Tb}), we have
\begin{equation}\label{eq:T3}
T^i \bab{\alpha^j}
=\bab{\alpha^{i+j}}
\end{equation}
for $i,j\in\intint{0}{\dmn^\degF -2}$. 
This implies
\begin{equation}\label{eq:T}
T^i = \begin{bmatrix} | & & | \\
                     \alpha^i & \cdots & \alpha^{i+\degF-1}\\
                      | & & | 
       \end{bmatrix} , \quad i\in\intint{0}{\dmn^\degF -2} 
\end{equation}
and hence,
\begin{equation}\label{eq:T1}
T^iT^j=T^{i+j}
\end{equation} 
and
\begin{equation}\label{eq:T2}
T^i+T^j=T^{\intk}
\end{equation} 
with $\intk$ satisfying
$\alpha^i+\alpha^j=\alpha^{\intk}$.  

To sum up, the map defined by
\[
\MatFa: \alpha^i \mapsto T^i,\quad i\in\intint{0}{\dmn^\degF -2},
\]
and $\MatFa(0)=O_{\degF}$ (zero matrix) is a homomorphism by (\ref{eq:T1})
and (\ref{eq:T2}).
Namely, (\ref{eq:Tbasic2}) holds.
Moreover, by (\ref{eq:T3}), 
%for $\xi,\xi'\in\myFkgen$, %4SPACE
for any $\xi,\xi'\in\myFkgen$, 
\begin{equation}\label{eq:T3alt}
\MatFa(\xi) \varphi_{\bsa}(\xi')\transp = \varphi_{\bsa}(\xi\xi')\transp.
\end{equation}

\subsection{Dual Bases \label{ss:dualbasis}}

In what follows, $\trace$ will be abbreviated as ${\rm Tr}$.
Put
\begin{equation}
\mypsi(\xi)=(\tracenos \xi, \tracenos \alpha \xi, \dots, 
%\varphi_{\bsa'}(\xi)=(\tracenos \xi, \tracenos \alpha \xi, \dots, 
\tracenos \alpha^{k-1} \xi ). \label{eq:dual_vec_rep}
\end{equation}
Then, it follows
%$\phi_{\bsa'}$ 
\begin{equation}\label{eq:T3alt_dual}
\mypsi(\xi) \MatFa(\xi')  = \mypsi(\xi\xi')
\end{equation}
for any $\xi,\xi'\in\myFkgen$.

{\em Proof of (\ref{eq:T3alt_dual}).}\/
We have
\begin{eqnarray*}
\lefteqn{\mypsi(\alpha^i)T }\\
&=& \tracenos \alpha^i (0,\ldots,0, f_0) \\
&&\mbox{}+ \tracenos \alpha^{i+1} (1,0,\ldots,0, f_1)
+\cdots\\
&&\mbox{}+ \tracenos \alpha^{i+k-1} (0,\ldots,0,1, f_{k-1}) \\
&=& (\tracenos \alpha^{i+1}, \dots, \tracenos \alpha^{i+k-1},x),
\end{eqnarray*}
where
\begin{eqnarray*}
x&=&\tracenos(\alpha^if_0+\cdots + \alpha^{i+k-1}f_{k-1})\\
&=&\tracenos\alpha^i(f_0+\cdots + \alpha^{k-1}f_{k-1})\\
&=&\tracenos\alpha^{i+k}.
\end{eqnarray*}
Hence, 
\begin{equation}\label{eq:Tb_dual}
\mypsi(\alpha^i) T =\mypsi(\alpha^{i+1}),
\end{equation}
which is the basic property that parallels (\ref{eq:Tb}).
Applying (\ref{eq:Tb_dual}) repeatedly, we obtain (\ref{eq:T3alt_dual}).
\enproof

It is well-known that any basis has a dual basis~\cite{LidlNied}.
In particular, denoting by $\bsa'$ the dual basis of  $\bsa$,
we have $\phi'=\varphi_{\bsa'}$ from (\ref{eq:dual_vec_rep}).%
\footnote{%
For the sake of self-containedness,
we remark that the existence of
a dual basis of $\bsa$ can be proved easily 
with the developments in this section as will be sketched.
%that $\mypsi$ is one-to-one.% This implies 
Using (\ref{eq:Tbasic2})
and (\ref{eq:T3alt_dual}), we can show
$\mypsi(\alpha^i)$ ranges over
all non-zero vectors in $\myFpower{k}$ as $i$ runs through
$\intint{0}{\dmn^k-2}$.
Hence, letting $j_i \in \intint{0}{\dmn^k-2}$ denote the number
such that 
$
\mypsi(\alpha^{j_i})
=(0,\dots,0,1,0,\dots,0),
$
where the $i$-th coordinate has the only non-vanishing component $1$,
we conclude that $\bsa'=(\alpha^{j_i})_{i=1}^{k}$
is the dual basis of $\bsa$ by (\ref{eq:dual_vec_rep}).}
Then, we can write
(\ref{eq:T3alt_dual}) as
\begin{equation}\label{eq:T3alt_dual2}
\varphi_{\bsa'}(\xi) \MatFa(\xi')  = \varphi_{\bsa'}(\xi\xi'),
\end{equation}
which makes good dual properties with (\ref{eq:T3alt}).

Thus, we have (\ref{eq:Tbasic}), which consists of
(\ref{eq:T3alt}) and (\ref{eq:T3alt_dual2}).
Since we have already shown (\ref{eq:Tbasic2}), the proof is complete.

\section{Syndrome Decoding for Concatenated Codes of CSS Type \label{ss:synd_dec}}

Having found a useful pair of dual bases $\bsa$ and $\bsa'$,
we set $\bsb=\bsa$ and $\bsb'=\bsa$ in this section.
We put
$\varphi=\varphi_{\bsa}$, $\varphi'=\varphi_{\bsa'}$
and $\MatF=\MatFa$ for simplicity.

\subsection{Decoding of $\dmn$-ary Images of Codes \label{ss:synd_n_k}}

We first recall how we can obtain a parity check matrix over $\myF$
of the `$\dmn$-ary image' of a code over an extension field $\myFk$.
We need some notation.
We extend the domain of $\varphi$ [$\varphi'$] to $\myFkpower{M}$,
where $M$ is a positive integer, in the natural manner:
We apply $\varphi$ [$\varphi'$] to each symbol of a word $x\in\myFkpower{M}$,
and denote
the resulting $kM$-dimensional vector over $\myF$ by $\varphi(x)$
[$\varphi'(x)$]. 
In the present case,
the $\dmn$-ary image of an $[N, \varKgen]$ linear code $D$ over $\myFk$
denotes the $[kN, k\varKgen]$ linear code 
$\varphi(D)$ or $\varphi'(D)$ over $\myF$. 

Let $H$ be a %an $(N-K)\times N$ 
parity check matrix of $D_1$.
We will show that we can find a matrix $H'$ such that
\begin{equation}\label{eq:H}
\varphi(xH\transp)=\varphi(x)H'\transp, \quad x\in\myFkpower{N}.
\end{equation}
Let us write $H=[h_{ji}]$ with $h_{ji}\in\myFk$.
Then, (\ref{eq:H}) holds for the matrix $H'=[\MatF(h_{ji})]$
with $\MatF=\MatFa$ as in Lemma~\ref{lem:expansion}.
This is a direct consequence of the first equation of 
(\ref{eq:Tbasic}) of Lemma~\ref{lem:expansion},
which can be rewritten as 
$\varphi(\xi') \MatF(\xi) \transp  = \varphi(\xi\xi')$.
In particular, we have, for $H'=[\MatF(h_{ji})]$,
\begin{equation}\label{eq:pchgmt1}
\varphi(D_1)=\{ y\in\myFpower{kN} \mid yH'\transp =\zrvb \}.
\end{equation}
We remark that we do not have to find the dual basis $\bsa'=\bsb'$ of $\bsa=\bsb$
explicitly in constructing $H'$.
A parity check matrix of $\varphi'(D_2)$ can similarly be obtained.

\subsection{Syndromes of Concatenated Codes of CSS Type \label{ss:synd_dec_cccss}}

%
%In the case where $\bsb=\bsa$ (and hence $\bsb'=\bsa'$),
Now we finally see the procedure for constructing
$G'_{j,i}$ in (\ref{eq:H4cat}) from a parity check matrix 
$H$ of $D_1$, which was presented in Section~\ref{ss:parity_check_m} (Steps 1 and 2), is useful for decoding the concatenated code $L_1/L_2^{\perp}$
as promised.

In fact, with the parity check matrix in (\ref{eq:H4cat}) 
and $G'_{j,i}$ constructed by the procedure,
the latter half of the syndrome is the same as
$
\varphi(x) H'\transp
$
by (\ref{eq:gg'}), where $\varphi=\varphi_{\bsb}$.
Namely, for $G'=[G'_{ji}]$, 
\[
\pi_1(x)G'\transp=\varphi(x)H'\transp.
\]
Hence, known procedures to estimate the error pattern from the syndrome
for $D_1$ 
%(with the representation $\varphi$) 
can be used to decode $\embone(D_1)$.

%\section{Performance of Concatenated Conjugate Codes \label{ss:performance}}
%\section{Decoding Error Probability of Efficiently Decodable Concatenated Codes of CSS Type \label{ss:performance}}
%and Decoding Complexity 
\section{Proof of Theorem~\ref{th:performance} \label{ss:performance}}

%{\em Proof of Theorem~\ref{th:performance}.}\/ 
We will establish the bound
by evaluating the decoding error probabilities of the concatenation $(L_1,L_2)$ of $(C_1,C_2)$ and $(D_1,D_2)$ as 
described in Section~\ref{ss:ccc}.
In the concatenation,
we use the pair $(C_1,C_2)$ attaining the exponent $E(W_j,r_j)$
for inner codes, and
generalized Reed-Solomon codes for outer codes $D_j$ 
of dimensions $K_j$ ($j=1,2$).
We consider an asymptotic situation 
where both $N$ and $n$ go to $\infty$,
$R_j = K_j/N$ approaches a fixed rate $R_j^*$, and 
$r_j$ approaches a rate $r_j^*$ ($j=1,2$).
The decoding error probability $\Pej$ of $L_j/L_{\bar{j}}$
%, where $\bar{1}=2$ and $\bar{2}=1$,
is bounded by
\begin{eqnarray*}
\Pej &\le & \sum_{i=\vart}^N \chooses{N}{i} \Peinj^i (1-\Peinj)^{N-i}\\
 &\le & q^{\vart \log_{q} \Peinj + (N-\vart) \log_{q}(1-\Peinj) + N h(\vart/N) }
\end{eqnarray*}
where $h$ is the binary entropy function, and $\vart=\lfloor (N-K_j)/2 \rfloor+1$.
%, i,e., $\Peinj \le a_n \dmn^{-n E(\rcl)}$,
Then, we have
\begin{eqnarray*}
\frac{1}{\Noa} \log_q \Pej &\le& 
\frac{\vart}{N} \Big[-E(W_j,r_j)+ \frac{o(n)}{n} \Big]\\
&&\!\!\!\!\! \mbox{}+ \frac{1}{n}\frac{N-\vart}{N} \log_q(1-\Peinj) + \frac{1}{n} h(\vart/N)
\end{eqnarray*}
for $j=1,2$.
Hence, 
the decoding error probability $\Pej$ of the concatenated code
$L_j/L_{\bar{j}}^{\perp}$
satisfy
\[
\limsup_{\Noa\to\infty} - \frac{1}{\Noa} \log_q \Pej 
\ge \frac{1}{2} \min_{j\in \{1,2\} } (1-R_j^*) E(r_j^*).
\]
for $j=1,2$.
Thus, we have the error bound in the theorem.

The detailed procedures for decoding and 
constructions
of parity check matrices for (general) concatenated 
codes $(L_1,L_2)$ have been presented in Sections~\ref{ss:parity_check_m} through \ref{ss:synd_dec}.
Note that
$n$ is proportional to $k \approx \log_q N$
and therefore that even with exhaustive
syndrome decoding, the decoding complexity for inner codes 
is at most polynomial in $q^n$, 
which is still polynomial in $q^k \approx N$ or $\Noa=nN$.
Hence, the constructed codes $L_1/L_2^{\perp}$ and $L_2/L_1^{\perp}$ are polynomially decodable.
This completes the proof.
%\enproof

\section{Minimum Distance of Quantum Codes \label{ss:mindis_intro}}

\subsection{Polynomial Constructions of Quantum Codes}

We move on to treating the issue of polynomial-time constructions of encoders
of quantum error-correcting codes. 
In what follows, the measure of goodness is
the minimum distance of codes.

As already mentioned, this issue was first treated in \cite{AshikhminLT01}. 
One important ingredient of the code construction in \cite{AshikhminLT01} is
a sequence of polynomially constructible algebraic geometry (AG) codes. 
These codes attain the Tsfasman-Vl\u{a}du\c{t}-Zink (TVZ) bound,
and are built on
a deep theory of modular curves~\cite{TsfasmanVladut}.
Alternative polynomially constructible geometric Goppa 
codes (AG codes) that attain the TVZ bound
were recently found~\cite{ShumAKSD01}.
We use these codes~\cite{ShumAKSD01}
%in our first construction of codes in what follows. %SOME VERSION
in our constructions of codes in what follows.
(Those familiar with the original polynomially constructible codes attaining the TVZ bound~\cite{TsfasmanVladut} can use them instead.)
The code construction in \cite{ShumAKSD01} relies on the theory of (algebraic) function fields~\cite{Stichtenoth}, so that we will also use the %notation and 
terminology in \cite{Stichtenoth}.

\subsection{Metrics for Quotient Spaces \label{ss:metrics}}

To evaluate minimum distance,
we use the metric naturally induced in a quotient space~\cite{hamada05qc}.
We begin with reviewing this metric.
Suppose we have spaces of the form $\cV=\cZ/\Bsmall$, 
where $\Bsmall \le \cZ$ are finite additive groups.
Given a non-negative function $\wght$ on $\cZ$, a function $\sD$
on $\cZ\times \cZ$ 
defined by $\sD(x,y)=\wght(y-x)$ is a metric if $\wght$ satisfies
(i) triangle inequality $\wght(x+y) \le \wght(x)+\wght(y)$, $x,y\in\cZ$,
(ii) $\wght(x)=0$ if and only if $x$ is zero,
and (iii) $\wght(x)=\wght(-x)$.
We have the following lemma~\cite[Appendix, A.3]{hamada05qc}.

\begin{lemma}  \label{lem:metric_qs}
%\cite{hamada05qc}. 
Given a function $\wght$ on $\cZ$, 
define $\cwght(\cst{x})=\min_{x\in\cst{x}}W(x)$ for $\cst{x}\in\cZ/B$.
Then, whichever of properties (i), (ii) and (iii) $\wght$ has,
$\cwght$ inherits the same properties from $\wght$.
\end{lemma}

The easy proof omitted in \cite{hamada05qc} is included below.

{\em Proof of Lemma~\ref{lem:metric_qs}.}\/
Given $\cst{x},\cst{y}\in\cZ/\Bsmall$,
let $x$ and $y$ attain the minimum of 
$\min_{x\in\cst{x}}\wght(\cst{x})$ and that of
$\min_{y\in\cst{y}}\wght(\cst{y})$, respectively.
Then,
\begin{eqnarray*}
\cwght(\cst{x})+\cwght(\cst{y}) &= &\wght(x) + \wght(y)\\
&\ge & \wght(x+y)\\
&\ge & \min_{z\in\cst{x+y}}\wght(z)\\
&= & \cwght(\cst{x+y})
\end{eqnarray*}
where $\cst{x+y}=\cst{x}+\cst{y}\in\cZ/\Bsmall$. 
This prove the statement on (i). That on (ii) is trivial.
To see that on (iii), it is enough to notice that when $z$ runs through
$\cst{x}=x+B$, $-z$ runs through $-x-B=-x+B=-\cst{x}$. 
\enproof

The lemma is, of course, applicable to the
Hamming weight, denoted by $\HamW$,
on the direct sum $\myFpowernoarg{n}$ of $n$ copies of 
an additive group $\myFnoarg$.
Namely, the quotient space $\myFpowernoarg{n}/B$ is endowed with
the weight $\HamW_{B}$,
defined by $\HamW_{B}(\tilde{x}) = \min_{x\in\tilde{x}} \HamW(x)$
for $\tilde{x}\in\myFpowernoarg{n}/B$, and the distance
$\HamD_{B}(x,y)=\HamW_{B}(y-x)$.
The minimum distance of a quotient code $C/B$ is denoted by
$\HamD_{B}(C)$ and defined as follows:
\begin{eqnarray}
\HamD_{B}(C) &=& \min\{ \HamD_{B}(\tilde{x},\tilde{y}) \mid \tilde{x},\tilde{y} \in C/B,\, \tilde{x}\ne \tilde{y} \} \nonumber\\
&= & \min\{ \HamW_{B}(\tilde{x}) \mid \tilde{x} \in C/B,\, \tilde{x}\ne B \} \nonumber\\
%&= & \min\{ \HamW(x) \mid x \in C \setminus B \} \nonumber\\ %. \label{eq:dBC}
&= & \HamW(C \setminus B) \label{eq:dBC}
\end{eqnarray}
where, for $A \subset \myFpowernoarg{n}$,
\[
%\HamW(A) = \min_{x \in A}  \HamW(w).
\HamW(A) = \min\{ \HamW(x) \mid x \in A \}.
\]

The minimum distance of the symplectic code generated by
a matrix $\cG=[G_x\, G_z]$, regarded as the quotient code $\spn \cG/\spn \cH$,
 %, as described in Section~\ref{ss:qc_cc}, 
is
\[
\min\{ \HamW([u,v]) 
 \mid (u|v) \in \spn \cG \setminus \spn \cH \}
\]
where $\spn\cH$ is the $\sypaltnoarg$-dual of $\spn\cG$ as given in Section~\ref{ss:qc_cc},
$[u,v]$ denotes $\big( (u_1,v_1),\cdots,(u_{\Noa},v_{\Noa}) \big) 
\in \cX^{\Noa}$, $\cX=\myFxpower{2}$, 
for $u=(u_1,\dots,u_{\Noa})$ and $v=(v_1,\dots,v_{\Noa})\in\myFxpower{\Noa}$,
and $\HamW([u,v])$ is the number of $i$ with $(u_i,v_i)\ne (0,0)$.
In particular, if $\cH$ is as in (\ref{G1G2})
with $\spn H_j = C_j^{\perp}$ ($j=1,2$),
the minimum distance of the CSS code $\css{C_1}{C_2}$ is given by
\[
\min\{ \HamD_{C_2^{\perp}}(C_1), \HamD_{C_1^{\perp}}(C_2) \}.
\]
The minimum distance of the code pair $(C_1,C_2)$ is also defined to
be $\min\{ \HamD_{C_2^{\perp}}(C_1), \HamD_{C_1^{\perp}}(C_2) \}$.
An $[[n,k]]$ symplectic code of minimum distance $d$
is called an $[[n,k,d]]$ symplectic code.
Similarly, an $[[n,k,d]]$ CSS code (code pair) is an $[[n,k]]$ 
CSS code (code pair) of minimum distance $d$.
An $[[n,k, \ge d ]]$ symplectic code 
refers to an $[[n,k, d' ]]$ symplectic code with $d' \ge d$.

\section{Minimum Distance of Concatenated Codes \label{ss:mindis}}

We will evaluate the minimum distances of
$L_1/L_2^{\perp}$ and $L_2/L_1^{\perp}$ for
$\Cone=\embone(D_1)+\overline{C_2^{\perp}}$ 
and $\Ctwo=[\embone(D_2^{\kperp})+\overline{C_2^{\perp}}]^{\perp}
=\embtwo(D_2)+\overline{C_1^{\perp}}$
for the concatenated code pair
as in Section~\ref{ss:ccc}.
For most part, 
we describe the argument only for $L_1/L_2^{\perp}$, the other case
being obvious by symmetry.

Here, 
an underlying idea 
that has brought about the results of the present work is explained.
The point is that both $L_1$ and $L_2^{\perp}$ have
the subspace $\overline{C_2^{\perp}}$,
and we encode no information
into $\overline{C_2^{\perp}}$. 
Namely, 
we encode a message into a `code-coset' of the form 
$u+L_2^{\perp} \in L_1/L_2^{\perp}$, 
which can be written in the form $\bigcup_{v} (v+\overline{C_2^{\perp}})$
since we have $\overline{C_2^{\perp}} \le L_2^{\perp}$ $(\le L_1)$.
This means there is no harm in dealing with the quotient space 
$\myFpower{\Noa}/\overline{C_2^{\perp}}$,
where $\Noa=nN$, in place of $\myFpower{\Noa}$,
which is to be dealt with when the conventional concatenated codes
are in question.
This is possible because
the space $\myFpower{n}/C_2^{\perp}$ is endowed
with the weight $\HamW_{C_2^{\perp}}$ 
as described in Section~\ref{ss:metrics}.

\begin{lemma}\label{lem:basic}
The minimum distance of the quotient code
$\Cone/\Ctwo^{\perp}=
[\embone(D_1)+\overline{C_2^{\perp}}]/
[\embone(D_2^{\kperp})+\overline{C_2^{\perp}}]$
is $d_1\Dout$, where $d_1=\HamD_{C_2^{\perp}}(C_1\ssi)$ and 
$\Dout=\HamD_{D_2^{\perp}}(D_1)$.
The minimum distance of the quotient code
$\Ctwo/\Cone^{\perp}=
[\embtwo(D_2)+\overline{C_1^{\perp}}]/
[\embtwo(D_1^{\kperp})+\overline{C_1^{\perp}}]$
is $d_2\Doutout$, where $d_2=\HamD_{C_1^{\perp}}(C_2\ssi)$ and 
$\Doutout=\HamD_{D_1^{\perp}}(D_2)$.
\end{lemma}

\begin{corollary}
The minimum distance of 
$\css{\Cone}{\Ctwo}$ is $\min \{ d_1\Dout, d_2\Doutout \}$.
\end{corollary}

{\em Proof.}\/
By symmetry, it is enough to show the first statement of the lemma.
We see this easily working with $\HamD_{\overline{C_2^{\perp}}}$.
In fact, for any $x\in D_1\setminus D_2^{\perp}$, the Hamming weight
of $x\in\myFkpower{N}$ 
is not smaller than $\Dout$,
and the $i$-th symbol $x_i\in\myFk$ of $x$ is mapped to (a representative of) $\tilde{y_i}
\in C_1\ssi/C_2^{\perp}$ for any $1 \le i \le N$
by $\embone\ssi$.
%by $\tilde{\embone}\ssi$.
Since $\tilde{y_i} \ne C_2^{\perp}$ has Hamming weight not less than $d_1$,
%$\HamW_{C_2^{\perp} \tilde{y}$,
the minimum weight of $L_1/L_2^{\perp}$ is lower-bounded by $d_1\Dout$.
The minimum weight is, in fact, $d_1\Dout$ since
we can choose a word  $x\in D_1\setminus D_2^{\perp}$ of weight $\Dout$
and a coset $\tilde{y_i}\in C_1/C_2^{\perp}$ of weight $d_1$.
Hence, we have the assertion in the lemma.
The corollary is trivial.
\enproof

\section{Bound on Minimum Distance \label{ss:bound}}

\subsection{The Bound}

In this section, we will present codes that exceed 
those in \cite{AshikhminLT01,ChenLX01}
%or the non-CSS-type codes in \cite{AshikhminLT01,rmatsumoto02}
in minimum distance for a wide region.
Specifically, we will prove the following theorem.

\begin{theorem} \label{prop:gen}
Let a number $0 \le R \le 1$ be given.
There exists a sequence of polynomially constructible 
$[[\Noa\dblith,\Koa\dblith,\doa\dblith]]$
code pairs 
%of minimum distance $\doa\dblith$
that satisfies
\[
\liminf_{\inu\to\infty}\frac{\doa\dblith}{\Noa\dblith} 
\ge \sup 
\frac{d_1d_2}{n(d_1+d_2)} \Big( 1-2\gtN- \frac{n}{k} R \Big),
\]
$\lim_{\inu\to\infty}\Koa\dblith/\Noa\dblith=R$,
and $\lim_{\inu\to\infty} \Noa\dblith = \infty$.
Here, $\gtN=(q^{k/2}-1)^{-1}$, and
the supremum is taken over all $(n,k,d_1,d_2)$
such that an $[[n,k]]$ code pair $(C_1,C_2)$ 
exists, $d_1=\HamW(C_1\setminus C_2^{\perp})$, 
$d_2=\HamW(C_2\setminus C_1^{\perp})$, 
and $q^k$ is a square (of a power of a prime).
%, i.e., $q^k$ is an even power of a prime.
\end{theorem}

{\em Remark.}\/ 
The polynomial constructibility of the sequence of code pairs,
$\{ (L_1\dblith,L_2\dblith) \}$, is to be understood as
the existence of a polynomial algorithm to produce a generator matrix
$G\ith$ of $L_1\dblith$ whose first $\Noa\dblith-K_2\dblith$ 
rows span $L_{2,\inu}^{\perp}$ for each $\inu$ (cf.\ Fig.~\ref{fig:css}). 
Note such a generator matrix of $L_1\dblith$
can be converted into the generator matrix of $L_2\dblith$ whose 
first $\Noa\dblith-K_1\dblith$ rows span $L_{1,\inu}^{\perp}$ polynomially. 
(The conversion can be done
by calculating the inverse of an $\Noa\dblith\times \Noa\dblith$ matrix
involving $G_{\nu}$.
To see this, put $C_j=L_j\dblith$ in Fig.~\ref{fig:css},
$j=1,2$.)
\enproof

The above definition of constructibility 
is suitable both for applications to wiretap channels
and for those to quantum error correction.
The former applications would be detailed elsewhere.
Regarding quantum error correction,
note we can readily obtain parity
check matrices, $H_1$ and $H_2$, of $L_{1,\inu}$ and $L_{2,\inu}$ from $G\ith$
as above.
Note also that the so-called stabilizer of the corresponding quantum code
is equivalent to the matrix $\cH$ associated 
with $(H_1,H_2)$ as in (\ref{G1G2}),
and 
a polynomial-time encoder of the quantum code is
obtained from this stabilizer efficiently for $q$ even~\cite{CleveGottesman97}.
(Here, the complexity is measured in terms of elementary quantum gates,
similarly to \cite{AshikhminLT01}, for two-level quantum systems.)
In fact, this directly follows from \cite{CleveGottesman97} for $q=2$.
To see it for $q=2^m$, note
$2^m$-ary CSS codes can be converted into binary symplectic codes
by expanding elements of $\myFnoarg_{2^m}$ using
dual bases. This is another application of (the extreme case
of) the concatenation method.
(More generally, by \cite{AshikhminKnill00}, 
$2^m$-ary symplectic codes can be converted into binary symplectic codes.) 
Because for $p$ odd, no established complexity measure
for circuits consisting of $p$-level quantum systems is known to the author,
we will assume that $q$ is even when discussing polynomial
complexity of quantum codes over a Hilbert space in what follows.
(In the binary case, standard elementary gates can be found, e.g., in \cite[p.~73]{KitaevSV}.)

\subsection{Proof of Theorem~\ref{prop:gen}}

%\subsection{Geometric Goppa Codes for Outer Codes }
First, we describe geometric Goppa codes
which are used as outer codes. 
We use codes over $\myFk$, where 
$q^k=p^m$ with some $p$ prime and $m$ even,
obtained from function fields
of many rational places (places of degree one) as outer codes.
Specifically, we use a sequence of
function fields $F\ith/\myFk$, $\inu=1,2,\ldots$, 
having genera $g\ith$ and at least
$N\ith+1$ rational places such that~\cite{%GarciaStichtenoth95AS1,
GarciaStichtenoth96AS2}
\begin{equation}\label{eq:DrVI}
%\lim_{\inu\to\infty} \frac{N\ith}{g\ith}=q^{k/2}-1.
\lim_{\inu\to\infty} \frac{g\ith}{N\ith}=\gtN \defeq \frac{1}{q^{k/2}-1}.
\end{equation}
(The resulting codes of length $N\ith$ are said to attain the TVZ bound.)
We put $\dvsr\ith=P_1+\cdots +P_{N\ith}$, where
$P_i$ are distinct rational places in $F\ith/\myFk$.
Let $G_2\dblith$ be a divisor of $F\ith/\myFk$
having the form $G_2\dblith=m_2 P_{\infty}$, $m_2 < N\ith$,
where $P_{\infty}$ is a rational place other than 
$P_1,\ldots,P_{N\ith}$.
%($\supp G\ith \cap \supp \dvsr\ith = \emptyset$),
Then, we have an $[N\ith,K_2\dblith]$ code of minimum distance $d''$,
where $K_2\dblith \ge \deg G_2\dblith +1 -g\ith$ and
$d'' \ge N\ith-\deg G_2\dblith$.
We use this code as outer code $D_2$, and 
let $D_1^{\perp}$ have a similar form.
Specifically, we put
\[
D_2 = C_{\cL}(\dvsr\ith,G_2\dblith)
\]
and
\[
D_1 = C_{\cL}(\dvsr\ith,G_1\dblith)^{\perp},
\]
where $G_1\dblith=m_1 P_{\infty}$ for some integer $m_1$,
and
\begin{equation}\label{eq:GGC}
C_{\cL}(\dvsr\ith,G) = \big\{ \big(f(P_1),\ldots,f(P_{N\ith})\big) \mid f \in \cL(G) \big\}.
\end{equation}
Here, $\cL(G)=\{ x \in F\ith \mid (x) \ge -G \}\cup\{ 0 \}$,
and $(x)$ denotes the (principal) divisor of $x$
(e.g., as in \cite[p.~16]{Stichtenoth}).
We require
\[
G_1\dblith \le G_2\dblith
\]
so that the CSS constraint $D_1^{\perp} \le D_2$ is fulfilled.

We also require
\begin{equation}\label{eq:degG}
%2g\ith -2 < \deg G_j\dblith =m_j < N\ith, \quad j=1,2.
2g\ith -2 < \deg G_j\dblith < N\ith, \quad j=1,2.
\end{equation}
Then, the dimension of $D_2$ is
\begin{equation}\label{eq:K1}
K_2\dblith=\dim G_2\dblith  = \deg G_2\dblith -g\ith+1
\end{equation}
and that of $D_1$ is
\begin{equation}\label{eq:K2}
K_1\dblith=N\ith-\dim G_1\dblith  = N\ith-\deg G_1\dblith +g\ith-1.
\end{equation}
The designed distance of $D_2$ is $N\ith-\deg G_2\dblith$,
and that of $D_1$ is $\deg G_1\dblith -2 g\ith +2$.
%which is a lower bound on the true minimum distance.

%\subsection{Proof of Theorem~\ref{prop:gen}}

With an inner $[[n,k]]$ code pair $(C_1,C_2)$ fixed, 
we consider an asymptotic situation 
where $K_j\dblith/N\ith$ approaches a fixed rate $R_j$
as $\inu$ goes to infinity ($j=1,2$).
Note that the limit of $[K_2\dblith-(N\ith-K_1\dblith)]/N\ith=(K_1\dblith+K_2\dblith-N\ith)/N\ith$,
the information rate of the outer quotient codes, is given by
\begin{equation}\label{eq:1}
\Rqt=R_1+R_2-1.
\end{equation}
Then, the overall rate of the concatenated code pair
$(L_1,L_2)$ has the limit
\begin{equation}\label{eq:Roa}
\Roa=\frac{k}{n}\lim_{\inu\to\infty}\frac{K_1\dblith+K_2\dblith-N\ith}{N\ith} = \frac{k}{n}\Rqt.
\end{equation}

If the quotient code $C_j/C_{\bar{j}}^{\perp}$,
where $\bar{1}=2$ and $\bar{2}=1$, 
has minimum distance not smaller than $d_j$, 
we can bound the minimum distance $\doaj$ 
of $L_j/L_{\bar{j}}^{\perp}$ using Lemma~\ref{lem:basic} as follows:
\begin{eqnarray}
\liminf_{\inu\to\infty}\frac{\doatwo}{\Noa\dblith}
&\ge & \frac{d_2}{n} \lim_{\inu\to\infty} \frac{N\ith - \deg G_2\dblith}{N\ith}  \nonumber\\
& = & \frac{d_2}{n} \lim_{\inu\to\infty} \Big(1-\frac{g\ith}{N\ith} - \frac{K_2\dblith}{N\ith}\Big) \nonumber\\
& = & \frac{d_2}{n} \lim_{\inu\to\infty} \Big(1-\frac{g\ith}{N\ith} - R_2\Big) \label{eq:rmd1}
\end{eqnarray}
by (\ref{eq:K1}), and 
\begin{eqnarray}
\liminf_{\inu\to\infty}\frac{\doaone}{\Noa\dblith}
&\ge & \frac{d_1}{n} \lim_{\inu\to\infty} \frac{\deg G_1\dblith-2g\ith}{N\ith}  \nonumber\\
& = & \frac{d_1}{n} \lim_{\inu\to\infty} \Big(1-\frac{g\ith}{N\ith} - R_1\Big)  \label{eq:rmd2}
\end{eqnarray}
by (\ref{eq:K2}).
Note the asymptotic form of (\ref{eq:degG}) is
\begin{equation}\label{eq:degGasympt}
\gtN
\le R_j
\le 1-\gtN, \quad j=1,2.
\end{equation}

It is expected that the best asymptotic bound will be 
obtained by requiring $d_1d' \approx d_2d''$, where $d'$ and $d''$ 
are the minimum distances of the outer codes as in Lemma~\ref{lem:basic}.
Thus, we equalize the bound in (\ref{eq:rmd1}) with
that in (\ref{eq:rmd2}), so that we have
\[ 
d_1 (1-\gtN - R_1)
=d_2( 1-\gtN - R_2).
\]
\begin{comment}
This, together with (\ref{eq:1}), gives
\[
R_1=\frac{(d_1-d_2)(1-\gtN) + d_2(\Rqt +1)}{d_1+d_2}.
\]
Using this and (\ref{eq:Roa}), 
\end{comment}
Using this, (\ref{eq:1}) and (\ref{eq:Roa}), 
we can rewrite (\ref{eq:rmd1}) and (\ref{eq:rmd2}) as
\begin{equation}\label{eq:gen}
%\liminf_{\inu\to\infty}\frac{\doa\dblith}{\Noa\dblith} 
\liminf_{\inu\to\infty}\frac{\doaj}{\Noa\dblith} 
\ge
\frac{d_1d_2}{n(d_1+d_2)} \Big( 1-2\gtN- \frac{n}{k} \Roa \Big)
\end{equation}
for $j=1,2$. 
%
%(The bound is negative for $q^k=4$.)

In the above construction, 
the second Garcia-Stichtenoth (GS) tower of function 
fields was used as $F\ith/\myFk$~\cite{%GarciaStichtenoth95AS1,
GarciaStichtenoth96AS2}.% 
\footnote{This tower is explicitly given by
$F_\nu=\myFk(x_1,\dots,x_\nu)$ with
$x_{\nu}^l+x_{\nu}=x_{\nu-1}^l/(x_{\nu-1}^{l-1}+1)$, $\nu=1,2,\dots$,
where $l=q^{k/2}$, and
$F_1=\myFk(x_1)$ with $x_1$ transcendental over $\myFk$.}
See \cite{ShumAKSD01} (also \cite{leonard01}) %,ShumPhD
for a polynomial algorithm to produce parity check matrices of
%one-point 
codes arising from the tower.
This, 
together with the method in Section~\ref{ss:parity_check_m},
gives needed parity check matrices of $L_1$ and $L_2$.
This completes the proof.

%We summarize the above argument in the following theorem.

\subsection{Calculable Bounds \label{ss:clb}}

First, we remark that 
Theorem~\ref{prop:gen} recovers the bound of \cite{ChenLX01}
by restricting the inner codes in the following manner.
Assume $C_1$ is an $[n=2t+1,k_1=2t,d_1=2]$ code such that
$C_1^{\perp} = \spn b_1$ with a fixed 
word $b_1\in(\myF\setminus\{0\})^n$, 
and $C_2$ is the $[n,k_2=2t+1,d_2=1]$ code, i.e., $\myFpower{n}$.
Then, the substitution of the inner code parameters
%into the bound 
into (\ref{eq:gen}) gives
the following bound~\cite{ChenLX01}:
%, which was claimed for $\dmn=2$ in 
\begin{equation}\label{eq:CLX}
l^{\rm CLX}_{t}(\Roa) =
%\frac{2}{3(2t+1)} \Big( 1-2\gtNt- \frac{2t+1}{2t} \Roa  \Big).
\frac{2}{3(2t+1)} \Big( 1-\frac{2}{\dmn^t-1} - \frac{2t+1}{2t} \Roa  \Big).
\end{equation}

When $q$ is a square, %an even power of a prime,
Theorem~\ref{prop:gen} also implies the following bound, which equals
the bound in \cite[Theorem~3.6]{FengLX06}.
%Hence, the bound in Theorem~\ref{prop:gen} is not worse than both (\ref{ChenLX}) and (\ref{eq:FengLX}).
Namely, if we put $n=k_1=k_2=d=1$ and $C_1=C_2=\myFpower{n}$, we have
\begin{equation}\label{eq:FengLX}
\liminf_{\inu\to\infty}\frac{\doa\dblith}{\Noa\dblith} 
\ge l^{\rm FLX}(R) \defeq
\frac{1}{2} \Big( 1 - \frac{2}{\sqrt{\dmn}-1} - R \Big).
\end{equation}
In particular, it was observed~\cite{FengLX06} that
the bound in (\ref{eq:FengLX}) exceeds
the Gilbert-Varshamov-type quantum bound 
%%$R= 1-2H_{q^2}(\delta)$
%$H_{q^2}^{-1}((1-R)/2)$
in some range for $\dmn \ge 19^2$
(as the Tsfasman-Vl\u{a}du\c{t}-Zink bound is larger than 
the classical Gilbert-Varshamov bound for $\dmn \ge 49$).
In \cite{FengLX06},
this bound was proved to be attained by quantum codes
described in a framework beyond symplectic codes;
it seems difficult to construct
encoders of polynomial complexity for their codes. 
By Theorem~\ref{prop:gen}, we have established that this bound
is attainable by polynomially constructible codes.

Thus,
the bound in Theorem~\ref{prop:gen} is not worse than the bounds 
in (\ref{eq:CLX}) and (\ref{eq:FengLX}).
We proceed to specifying an illustrative inner code pair, which
results in a significant improvement.

Take two (not necessarily distinct) 
words $b_1,b_2\in(\myF\setminus\{0\})^n$ 
and set $C_j^{\perp} = \spn b_j$, $j=1,2$.
We require the condition (\ref{eq:css_cond}), i.e.,  $\dpr{b_1}{b_2}=0$,
%Then, $C_j$ has minimum distance two, $j=1,2$.
and use the $[[n,n-2,2]]$ code pair $(C_1,C_2)$
%of minimum distance two 
as inner codes ($d_1=d_2=2$).
%Here, we assume that %$t > 0$ is an integer
%$q^{2t}$ is a square of a power of a prime.
With this choice of the inner code pair,
%for which $n=2t+2,k=2t,d_1=d_2=2$, %$[[2t+2,2t,2]]$ 
%Lemma~\ref{lem:main0} 
Theorem~\ref{prop:gen} immediately yields
the following proposition, where we put $t=k/2=(n-2)/2$.

\begin{proposition} \label{prop:main}
Let a number $0 \le R \le 1$ be given.
There exists a sequence of polynomially constructible 
$[[\Noa\dblith,\Koa\dblith,\doa\dblith]]$
code pairs 
%of minimum distance $\doa\dblith$
that satisfies
\[
\liminf_{\inu\to\infty}\frac{\doa\dblith}{\Noa\dblith} 
%\ge \sup_{t\in\SINT,\, t\ge 3}\frac{1}{t+1}\Big(1-\frac{1}{\dmn^{t}-1} - \frac{R(t+1)/t+1}{2}\Big),
\ge \sup \frac{1}{t+1}\Big(\frac{1}{2}-\frac{1}{\dmn^{t}-1} - \frac{t+1}{2t}R\Big),
\]
$\lim_{\inu\to\infty}\Koa\dblith/\Noa\dblith=R$,
and $\lim_{\inu\to\infty} \Noa\dblith = \infty$.
Here, the supremum is taken over $t$ such that $\dmn^{t} \ge 3$ 
is a power of a prime.
\end{proposition}

\begin{comment}
For example, if $q=2$, $n$ is necessarily even by $\dpr{b_1}{b_2}=0$,
$C_1^{\perp}=C_2^{\perp}=\spn (1,\cdots,1) \le C_1$, where $1^n=(1,\ldots,1)\in\myFpower{n}$
and hence, $C_1=C_2$ is 
composed of all even-weight words in $\myFpower{n}$.
\end{comment}

\section{Comparisons \label{ss:cmp}}

\begin{figure} %[h]
\begin{center}
%FIGURE1
\includegraphics[scale=0.55]{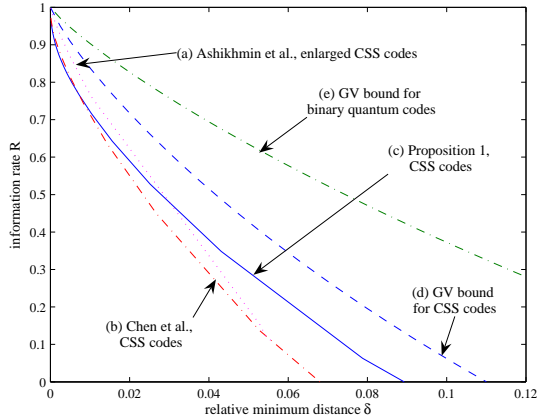} % Here is how to import EPS art
%\vspace*{-2ex}
\caption{Bounds on the minimum distance of binary CSS and enlarged CSS codes.
The plotted bounds are
(a) bound attainable by enlarged CSS codes in \protect\cite{AshikhminLT01},
(b) the bound attainable by the CSS codes
in \protect\cite{ChenLX01},
(c) the improved bound on the minimum distance of 
CSS codes in Proposition~\protect\ref{prop:main},
(d) the Gilbert-Varshamov-type bound $R=1-2H_2(\delta)$ for CSS codes~\protect\cite{CalderbankShor96}, where $H_2(x)=-x\log_2 x-(1-x)\log_2(1-x)$,
and 
the Gilbert-Varshamov-type bound $R=1-H_2(\delta)-\delta\log_2 3$ for binary
quantum codes~\protect\cite{crss97}.
These codes are polynomially constructible except (d) and (e).
\label{fig:binary}}
\end{center}
\end{figure}

In this section, we will %first
compare the bound in Proposition~\ref{prop:main}
with the best bounds known in the binary case ($\dmn=2$).
%the bound in \cite{ChenLX01} for CSS codes 
Let a point $(\delta,R)$ be called attainable
if we have a sequence of polynomially constructible
$[[N\ith,K\ith,d\ith]]$ CSS codes $\css{C_1\dblith}{C_2\dblith}$
%such that $\limsup_\inu d\ith/N\ith \ge \delta$,
such that $\liminf_\inu d\ith/N\ith \ge \delta$,
$\liminf_\inu K\ith/N\ith \ge R$, and $\lim_{\inu} N\ith =\infty$.
Then, by Proposition~\ref{prop:main}, the points in 
$
\bigcup_{t \ge 3} \cM_t
$
is attainable,
where 
\begin{equation}\label{eq:reagion}
\cM_t = \{ (\delta,R) \mid \mbox{$0 \le \delta \le 1$ and 
$0 \le R \le R_t(\delta)$} \}
\end{equation}
and
\begin{equation}\label{eq:R_t}
R_t(\delta) =\frac{t}{t+1}\Big(1-\frac{2}{q^t-1}\Big) -2 t \delta.
\end{equation}
Note $R=R_t(\delta)$ is merely a rewriting of
\[
\delta = l_{t}(R) \defeq \frac{1}{t+1}\Big(\frac{1}{2}-\frac{1}{\dmn^{t}-1} - \frac{t+1}{2t}R\Big).
\]
Hence, our bound is the upper boundary of the region
$
\bigcup_{t \ge 3} \cM_t,
$
which
is the envelope formed by the collection of the straight lines $R=R_t(\delta)$,
$t \ge 3$.
This bound, together with previously known polynomial bounds, is plotted in Fig.~\ref{fig:binary}.
The improvement is clear from the figure.

\section{Steane's Enlargement of CSS Codes \label{ss:steane}}

\subsection{Effect of General Inner Codes and Another Effect}

Our concatenation method is applicable to any inner CSS codes.
It is this flexibility that has brought about the improvement
as presented in Fig.~\ref{fig:binary}.
From the figure, however, 
one sees the bound in \cite{AshikhminLT01} retains
the superiority in some region,
which must come from a distinct nature of the code construction of \cite{AshikhminLT01},
namely, the property of enlarged CSS codes~\cite{steane99e}.
In this section, we present another construction of codes
which has both the merits of the flexibility of inner codes
and the good distance property of enlarged CSS codes.

\subsection{Enlarged CSS Codes}

Enlarged CSS codes are a class of quantum error-correcting
codes proposed by Steane~\cite{steane99e}. 
These can be viewed as enlargements
of CSS codes $\css{L_1}{L_1}$
%corresponding to the code pairs $(L_1,L_2)$ with $L_1=L_2$,
and are defined as follows.
The definition below is general in that it applies to
any prime power $\dmn$.
%In this section, we use the notation in \cite{steane99e}.

Assume we have an $[\Noa,\Koa]$ linear code $\genC$ which contains
its dual, $\genC^{\perp} \le \genC$, and which can be enlarged
to an $[\Noa,\Koa']$ linear code $\genC'$. 
%, where $\Koa' \ge \Koa+\dmn$.
Let a generator matrix $W$ of $\genC'$ has the form
\begin{equation}\label{eq:Gpr}
\matGpr=\left[ \begin{array}{c} 
U\\
V
\end{array}
\right]
\end{equation}
where $U$ and $V$ are of full rank, and $U$ is a generator matrix
of $\genC$, and let $\stA$ be a
$(\Koa'-\Koa) \times (\Koa'-\Koa)$ invertible matrix.
Then, the code generated by 
\begin{equation}\label{eq:stG}
{\cal G} = \left[ \begin{array}{c|c} 
U   & 0 \\
  0 & U \\
V   & \stA V  \end{array}
\right]
\end{equation}
is a symplectic code~\cite{steane99e}.
We denote this code by $\stn(\matGpr,\stA)$.
\begin{comment}
(Formally, we allow $\stA$ to be `$0\times 0$ matrix.'
In this case, $\genC=\genC'$ and $\stn(\matGpr,\stA)$ is the CSS code $\css{\genC}{\genC}$.)
%corresponding the code pair $(C,C)$.)
\end{comment}

Now suppose that $xM \ne \lambda x$ for any $\lambda \in \myF$, i.e.,
that $\stA$ is fixed-point-free when it acts %, from right, 
on
the projective space $(\myFpower{\Koa'-\Koa}\setminus\{ \zrv \})/\sim$,
where $\zrv$ denotes the zero vector and
$x\sim y$ if and only if $y=\lambda x$ for some $\lambda\in\myF$.
This is possible by Lemmas~\ref{lem:steaneAexist} and \ref{lem:steaneAexist2} in Appendix~\ref{app:enl}
if the size $\Koa '-\Koa$ of $M$ is not less than $2$.
Such a choice of $\stA$ results in a good symplectic code as
the next lemma and corollaries show. 
These are essentially from
\cite{steane99e} and \cite{CohenEL99}.
%or a more general form in \cite{CohenEL99}.

\begin{lemma}\label{lem:steane}
Assume we have an $[\Noa,\Koa]$ linear code $\genC$ which contains
its dual, $\genC^{\perp} \le \genC$, and which can be enlarged
to an $[\Noa,\Koa']$ linear code $\genC'$, where $\Koa' \ge \Koa+2$.
Take a full-rank generator matrix $W$ of $\genC'$
having the form in (\ref{eq:Gpr}),
where $U$ is a generator matrix of $\genC$, 
and a fixed-point-free matrix $M$.
Then, $\stn(\matGpr,\stA)$ is an 
$[[\Noa,\Koa+\Koa'-\Noa,\ge\!\min\{ d, d'' \} ]]$
symplectic code, where
$d = \HamW(\genC \setminus \genC'\mbox{}^{\perp})$ and
\[
d'' = 
\min \{ \HamW([u,v]) \mid u,v \in \genC' \setminus \genC'\mbox{}^{\perp}, \
\forall \lambda \in\myF,\, v \ne  \lambda u  \}.
\]
\mbox{}
\end{lemma}

\begin{corollary}\label{coro:steane_CEL}
Under the assumptions of the lemma, $\stn(\matGpr,\stA)$ is an 
$[[\Noa,\Koa+\Koa'-\Noa,\ge\!\min\{ d, d_2' \} ]]$
symplectic code, where
\[
d_2' = 
\min \{ \HamW([u,v]) \mid u,v \in \genC' \setminus \{ \zrv \}, \
\forall \lambda \in\myF,\, v \ne  \lambda u  \}.
\]
\mbox{}
%with $\zrv$ denoting the zero vector.
\end{corollary}

\begin{corollary}\label{coro:steane}
Under the assumptions of the lemma, $\stn(\matGpr,\stA)$ is an 
$[[\Noa,\Koa+\Koa'-\Noa,\ge\!\min\{ d, \lceil \frac{\dmn+1}{\dmn} d' \rceil \} ]]$ symplectic code, 
where $d' = \HamW(\genC' \setminus \genC'\mbox{}^{\perp})$.
\end{corollary}

{\em Remarks.}\/
The premise of the lemma implies
\begin{equation}\label{eq:tower}
\genC'\mbox{}^{\perp} \le  \genC^{\perp} \le \genC \le \genC'.
\end{equation}
% tower of codes
In Steane's original bound~\cite[Theorem~1]{steane99e},
$\HamW(\genC \setminus \{ \zrv \})$ and
$\HamW(\genC' \setminus \{ \zrv \})$ were used in place of %our
%$d$ and $d'$,
$d=\HamW(\genC \setminus \genC'\mbox{}^{\perp})$ and
$d'=\HamW(\genC' \setminus \genC'\mbox{}^{\perp})$,
respectively.

The quantity $d_2'$
is the second generalized Hamming weight of $\genC'$.
Corollary~\ref{coro:steane_CEL} with $q=2$
was given in \cite{CohenEL99} 
to improve significantly on the bound in \cite{steane99e}.
%i.e., Corollary~\ref{coro:steane} with $q=2$.
%In what follows, we only need Corollary~\ref{coro:steane}.
\mbox{} \enproof %LAYOUT

To prove Lemma~\ref{lem:steane} and corollaries, we should only
examine the proof of Theorem~1 in \cite{steane99e} 
or the proof of its refinement, Theorem~2 of \cite{CohenEL99},
noting that we may assume $H'$, the generator matrix of
$\genC'\mbox{}^{\perp}$, is a submatrix of $U$ ($G$ in \cite{steane99e}).
In particular, if $q=2$, this can be done without pain.
A proof for the general prime power $q$ is included in Appendix~\ref{app:enl}.

\subsection{Enlargement of Concatenated Codes of the CSS Type}

In \cite{AshikhminLT01}, Steane's construction was applied
to binary images of geometric Goppa codes $D^{\perp} \le D \le D'$.
The binary image of a code $D_1$ over $\myFk$ denotes $\embone(D_1)$
with $n=k$, $q=2$ in the notation of Section~\ref{ss:ccc}.
We can regard the codes in \cite{AshikhminLT01}
the enlargement of $\big(\embone(D_1), \embtwo(D_2)\big)$
with $\embone=\embtwo$ and $D_1=D_2$, i.e.,
$\big(\embone(D_1), \embone(D_1) \big)$,
where the inner code pair $(C_1,C_1)=(\myFpower{k},\myFpower{k})$
is the trivial $[[n,n]]$ code.

In what follows, we establish
a similar bound attained by
some enlargement of $\big(\embone(D_1), \embone(D_1)\big)$
with a geometric Goppa code $D_1$ 
%with $\embone=\embtwo$
in the case where an $[[n,k]]$ inner code pair $(C_1,C_1)$ is not
necessarily $(\myFk,\myFk)$.
%the $[[n,n]]$ inner code pair
In our construction, 
we also need the concatenation method of Section~\ref{ss:ccc},
so that we retain the notation therein.
We require the existence of $C_1$ satisfying the following conditions 
in order to make $\embone$ and $\embtwo$ equal to each other.

{\em Conditions}.\/
\begin{enumerate} %{condition} \label{cond:1}
\item[(A)]  $C_1^{\perp} \le C_1 \le \myFpower{n}$.
\item[(B)] We have vectors $\myg_j$, $j=\intint{1}{k}$, 
which  satisfy $\dpr{\myg_i}{\myg_j} = \delta_{ij}$
and which, together with a basis of $C_1^{\perp}$,
form a basis of $C_1$, 
where $k= 2 \dim_{\myF} C_1 - n$.
\item[(C)]
$\myFk$ has a self-dual basis $\big(\mybeta_j\big)_{j=1}^{k}$.
%for the above $k$.
\end{enumerate} %condition}

Note (A), together with $k=2 \dim_{\myF} C_1 - n$, implies that 
$(C_1,C_1)$ is an $[[n,k]]$ code pair, cf.~(\ref{eq:css_cond}) and
(\ref{eq:css_k}).
Recall we have required
$\trace \mybeta_i \mybetapr_j = \dpr{\myg_i}{\mygpr_j} = \delta_{ij}$
%, where $\delta_{ij}=1$ if $i=j$ and $\delta_{ij}=0$ otherwise,
in constructing the map $\pi_m: \beta_j^{m)} \mapsto g_j^{m)}$,
$j\in\intint{1}{k}$, $m=1,2$ (Section~\ref{ss:ccc}).
Hence, under the conditions (A), (B) and (C), 
we have $\pi_1=\pi_2$ as desired by setting
\[
\big(\mybetapr_j\big)_{j=1}^{k} = \big(\mybeta_j\big)_{j=1}^k \quad \mbox{and}
\quad
\big(\mygpr_j\big)_{j=1}^{k} = \big(\myg_j\big)_{j=1}^k.
\]

Similarly to \cite{AshikhminLT01}, we use
a tower of codes $D^{\perp} \le D \le D'$ over $\myFk$,
all of which arise from some sequence of function fields
$F_1, F_2 , \cdots$,
such as given in \cite{GarciaStichtenoth96AS2}
and have the form $a\cdot C_{\cL}(\dvsr\ith,G)$, where
\[
C_{\cL}(\dvsr\ith,G) = \big\{ \big(f(P_1),\ldots,f(P_{N})\big) \mid f \in \cL(G) \big\}
\]
and
\[
a\cdot D = \{ (a_1x_1,\dots,a_N x_N) \mid
(x_1,\dots,x_N) \in D \}
\]
for some $a=(a_1,\dots,a_N)\in (\myF\setminus \{ 0 \})^N$.
Specifically, 
\[
%D^{\perp} =C_{\cL}(\dvsr\ith,G)^{\perp}, \ %\quad \!
D = a \cdot C_{\cL}(\dvsr\ith,G), \quad %\ %\quad \!
D' = a \cdot C_{\cL}(\dvsr\ith,G'),
\]
where $\dvsr\ith=P_1+\cdots +P_{N}$,
$P_i$ are distinct rational places in $F\ith/\myFk$,
and $G,G'$ are divisors of $F\ith/\myFk$
whose supports are disjoint with that of $\dvsr\ith$.
Put $\lim_{\inu} g\ith/N =\gtNhat$.
%\begin{equation}\label{eq:gtNhat}
%\lim_{\inu\to\infty} \frac{g\ith}{N}=\gtNhat .
%\end{equation}
A major difficulty of the construction resides in
the constraint $D^{\perp} \le D \le D'$, i.e.,
$G^{\perp} \le G \le G'$
when $D^{\perp}$ is written as $a \cdot C_{\cL}(\dvsr\ith,G^{\perp})$.

Under this condition, we apply Lemma~\ref{lem:steane}
putting $\genC=\embone(D)+\bar{C_1^{\perp}}$ and
$\genC'=\embone(D')+\bar{C_1^{\perp}}$, where $\embone$ and 
$\bar{C_1^{\perp}}$ are as in Section~\ref{ss:ccc}.
%NOTE: 

%Since the inner code pair $(C_1,C_2)$ is such that $C_1=C_2$, 
Since $C_1=C_2$, Theorem~\ref{th:duals_ccc} implies
$\genC^{\perp}=\embone(D^{\perp})+\bar{C_1^{\perp}}$ and
$\genC'\mbox{}^{\perp}=\embone(D'\mbox{}^{\perp})+\bar{C_1^{\perp}}$.
Namely, in the present case,
the tower in (\ref{eq:tower}) can be written as
\begin{equation}\label{eq:tower2}
\embone(D'\mbox{}^{\perp}) + \cdB
\le \embone(D^{\perp}) + \cdB \le \embone(D) + \cdB
\le \embone(D') + \cdB
\end{equation}
where $\cdB=
\overline{C_1^{\perp}}=\bigoplus_{i=1}^N C_1^{\perp}$.
Keeping in mind evaluating $\HamD_{\cdB}$, rather than $\HamD$, 
is enough for our purpose,
one can calculate the bound 
in a manner similar to that in \cite{AshikhminLT01}, 
which leads to the next proposition.
A proof may be found in Appendix~\ref{app:enl}.

\begin{proposition}\label{prop:enl}
Assume
we have an $[[n,k, d]]$ code pair $(C_1,C_1)$
over $\myF$ for which the conditions (A), (B) and (C) are true, 
a sequence of function fields $\{ F\ith/\myFk \}$, 
and a sequence of positive integers $\{ N\ith \}$
with $N\ith \to \infty$ ($\inu \to \infty$)
satisfying the following three conditions for any $R' > R \ge 1/2$.
(i) For all large enough $\inu$, we have $N=N\ith$ distinct rational places 
$P_1,\cdots, P_{N}$ in $F\ith/\myFk$, and
divisors $G=G\ith$ and $G'=G'\ith$ of $F\ith/\myFk$ such that
(a) the supports of $G,G'$ contain none of $P_1,\cdots, P_{N}$,
(b) $G\le G'$, and (c) $D^{\perp} \le D$ for $D= a \cdot C_{\cL}(\dvsr,G)$
with some $a=(a_1,\dots,a_N)\in (\myF\setminus \{ 0 \})^N$,
where $\dvsr=P_1+\cdots+P_{N}$.
(ii) The genus $g\ith$ of $F\ith/\myFk$ satisfies
\[
\gtNhat \defeq \lim_{\inu\to\infty} \frac{g\ith}{N} < \frac{1}{2}.
\]
(iii) $G$ and $G'$ fulfill
\begin{equation*}
\lim_{\nu\to\infty} \frac{\deg G - g\ith}{N} \ge R, \quad
\lim_{\nu\to\infty} \frac{\deg G' - g\ith}{N} \ge R'.
\end{equation*}
Then,
we have a sequence of
$[[\Noa,\Koa'',\doa]]$ symplectic codes $\stn(W\ith,M\ith)$
that satisfies $\lim_{\inu} \Noa = \infty$,
\begin{equation*} %\label{eq:rateS2}
\liminf_{\nu\to\infty} \frac{\Koa''}{\Noa} \ge \Roa
\end{equation*}
and
\begin{equation*} %\label{eq:boundSgen}
\liminf_{\nu\to\infty} \frac{\doa}{\Noa} \ge 
%l^{\rm enl}_t \defeq  
\frac{(\dmn+1)d}{(2\dmn+1)n}\Big( 1 -2\gtNhat -\frac{n}{k} \Roa \Big)
\end{equation*}
for any rate
\begin{equation*} %\label{eq:Rgehalf}
\Roa \ge \frac{k}{2(\dmn+1)n}(1-2\gtNhat).
\end{equation*}
\mbox{}
\end{proposition}

{\em Remark.}\/
The assumption that for any $R' > R \ge 1/2$, (iii) holds says
$\deg G$ and $\deg G'$ are flexible enough
%($\dim D \ge \deg G-g\ith+1$,
($R \ge 1/2$ stems from $D^{\perp} \le D$).
%; recall $\dim D \ge \deg G-g\ith+1$).
This, as well as the other two, is fulfilled for some 
\begin{equation}\label{eq:gALT}
\gtNhat \le \gtN/(1-\gtN) = (\gtN^{-1}-1)^{-1},
\end{equation}
where $\gtN=(\dmn^{k/2}-1)^{-1}$,
and for polynomially constructible codes $D$ and $D'$, $D^{\perp}\le D \le D'$,
if $q^{k/2}$ is even~\cite{AshikhminLT01}.
Namely, in \cite{AshikhminLT01}, they showed how such $D$ and $D'$ with
(\ref{eq:gALT}) can be obtained
from general geometric Goppa codes attaining the TVZ bound.
%($\gtNhat = \gtN$).
If the codes from \cite{VossHoholdt97,rmatsumoto02,Stichtenoth06}
are used instead,
the premise of the proposition is true for $\gtNhat = \gtN$.
However, we should emphasize that
using the suboptimal value $\gtNhat= (\gtN^{-1}-1)^{-1}$ 
in \cite{AshikhminLT01} is to
establish the polynomial constructibility of the codes.
We remark that their argument to obtain codes with 
$\gtNhat= (\gtN^{-1}-1)^{-1}$ 
(see Theorem~4 of \cite{AshikhminLT01}), 
is applicable to 
general geometric Goppa codes including the one that has been used in 
this paper, i.e., the code in \cite{ShumAKSD01}.%
\footnote{The status of results along the lines of
\cite{VossHoholdt97,rmatsumoto02,Stichtenoth06} is as follows. 
Though the codes in \cite{VossHoholdt97,rmatsumoto02,Stichtenoth06}
have the desirable properties  $D^{\perp} \le D$ and $\gtNhat = \gtN$, 
they have not been proved to be polynomially constructible.
It is true that the descriptions of these codes in the form $C_{\cL}(\dvsr,G)$
are explicit, i.e., the underlying sequence of function fields
and %divisors 
$\dvsr,G$ have been specified explicitly.
However, we need to solve an additional problem of
finding generator matrices of $D=a \cdot C_{\cL}(\dvsr,G)$ 
and $D'= a \cdot C_{\cL}(\dvsr,G')$ to establish
the polynomial constructibility of $D$ and $D'$.
%$C_{\cL}(\dvsr,G)$ or $C_{\cL}(\dvsr,G)^{\perp}$.
The problem of constructing optimal codes $D$,
which arise from explicit function fields~\cite{GarciaStichtenoth96AS2},
in polynomial time
{\em without the constraint $D^{\perp} \le D$}\/
had attracted interest until it was solved in \cite{ShumAKSD01}.}
As remarked in \cite{AshikhminLT01}, the necessity to construct
codes with $D^{\perp} \le D$ has never arisen before \cite{AshikhminLT01}.
\begin{comment}
Note we did not need to place the restriction $D^{\perp} \le D$ in Section~\ref{ss:bound}. 
\end{comment}
\enproof

This proposition recovers the bound in \cite{AshikhminLT01} by putting
$\gtNhat = (\gtN^{-1}-1)^{-1}$,
$\dmn=2$, $n=k=2m$ and $d=1$.
As in Section~\ref{ss:bound}, we take inner code pairs
with minimum distance two as an example.
\begin{lemma}\label{lem:dis2}
For any square $\dmn$ of a power of two,  and $n \ge 3$,
we have an $[n,n-1]$ linear code $C_1$ over $\myF$ of the following properties.
(A') $C_1^{\perp}=\spn b$
for some vector $b \in(\myF\setminus\{0\})^n$ with $\dpr{b}{b}=0$.
(B') We have vectors $\myg_j$, $j\in\intint{1}{n-2}$, 
which  satisfy $\dpr{\myg_i}{\myg_j} = \delta_{ij}$
and which, together with $b$, form a basis of $C_1$.
\end{lemma}

%The easy proof of Lemma~\ref{lem:dis2} is omitted.
A constructive proof of Lemma~\ref{lem:dis2} is included in Appendix~\ref{app:dis2}.
For $C_1$ in the lemma, $(C_1,C_1)$ is an $[[n,n-2,2]]$ code pair.
Recall the well-known fact that $\myFk$ has a self-dual basis over $\myF$ if
$q$ is even~\cite{SLempel80} (also \cite[p.~75]{LidlNied} for the statement only).
Thus, for a square of a power of two $\dmn=2^{2m}>2$ and $n=3,4,\dots$, we have $C_1$
that satisfy the conditions (A), (B) and (C).

For these parameters $q,n,k=n-2,d=2$
and $\gtNhat=\gtNhat(k)\defeq (\gtN^{-1}-1)^{-1}$,
the bound in Proposition~\ref{prop:enl} becomes
\iffalse
\[
\liminf_{\nu\to\infty} \frac{\Koa''}{\Noa} \ge \Roa
\]
and
\fi
\begin{equation}\label{eq:boundS}
\liminf_{\nu\to\infty} \frac{\doa}{\Noa} \ge 
%l^{\rm enl}_k \defeq   %quat
\frac{10}{9(k+2)}[ 1 -2\gtNhat(k) ] -\frac{10}{9k} \Roa
\end{equation}
where
\begin{equation}
\Roa \ge \frac{k}{10(k+2)}[1-2\gtNhat(k)],
\end{equation}
and this is attainable by polynomially
constructible $[[\Noa,\Koa'',\doa]]$ symplectic codes.

\subsection{Comparisons}

\begin{figure} %[h]
\begin{center}
%FIGURE1
\includegraphics[scale=0.55]{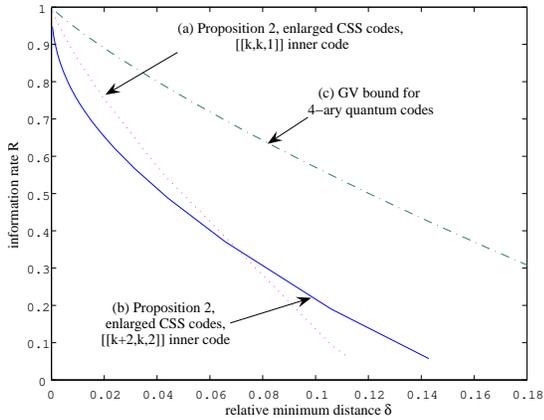} % Here is how to import EPS art
%\vspace*{-2ex}
\caption{Bounds on the minimum distance of quaternary quantum codes ($q=4$).
The plotted bounds are
(a) the bound on the minimum distance in Proposition~\protect\ref{prop:enl} with $n=k$ and $d=1$,
(b) the bound 
in Proposition~\protect\ref{prop:enl} with $k=n-2$ and $d=2$,
and (c) 
the Gilbert-Varshamov-type bound $R=1-H(x)-\delta\log_4 15$ for quaternary quantum codes~\protect\cite{AshikhminKnill00},
where $H(x)=-x\log_4(x)-(1-x)\log_4(1-x)$.
These codes are polynomially constructible except (c).
\label{fig:4ary}}
\end{center}
\end{figure}

%\subsubsection{Case $q=4$}
The constructive bound in (\ref{eq:boundS}),
as well as the similar bound with the $[[k,k,1]]$ inner code, 
is plotted in Fig.~\ref{fig:4ary} for $q=4$.
These bounds use constructible geometric Goppa codes with 
$\gtNhat \le (\gtN^{-1}-1)^{-1}$.
%
%
%The comparison in Section~\ref{ss:cmp} is on consructible codes for $\dmn=2$.
One sees that the enlargement of concatenated CSS codes
with the $[[k+2,k,2]]$ inner code pair outperforms
the enlargement with the $[[k,k,1]]$ inner code pair for relatively large $\delta$.
Namely, the flexibility of inner code pairs is effective
also for constructions of enlargements of concatenated CSS codes.%
\footnote{The author did not find any instance of the bound 
(\ref{eq:gen}), which uses CSS construction,
that exceed the bounds (a) and (b) in Fig.~\ref{fig:4ary} except 
Proposition~\ref{prop:main} with $t=2$.
This exceeds (a) and (b) slightly only in the narrow interval
$1/7\approx 0.1429 \le \delta \le 0.1444$, 
where the bounds (a) and (b) vanish.}

For any prime power $\dmn$, observe that
the bound in Proposition~\ref{prop:enl} with $n=k=d=1$
and $\gtNhat=\gtN$
exceeds the bound in (\ref{eq:FengLX}).
Thus, finding constructible dual-containing codes with $\gtNhat=\gtN$
would be an interesting future topic (cf.\ footnote 5).

%\section{Remark and Summary}
\section{Summary and Remarks \label{ss:summary}}
%\subsection{Constructibility}

%In summary, 

A method for concatenating quantum codes was presented.
We also showed how to construct parity check matrices
of concatenated quantum codes
preserving the syndromes for outer codes before concatenation. 
%retaining the structure of extension field over which the outer codes are built. 
Based on these results, it was proved that the so-called
Shannon rate is achievable by efficiently decodable codes.
The minimum distance of concatenated 
quantum codes was also evaluated 
to demonstrate that the proposed code class
contains codes superior to those previously known.

We remark that for the codes $L/B$ obtained by means of concatenation in this work,
the minimum distance
$\HamD_B(L)=\HamW(L\setminus B)$ of $L/B$ is significantly larger than 
the usual minimum distance $\HamW(L\setminus\{ \zrv \})$ of $L$.
In fact, $B$ contains the space of the form
$\bigoplus_{i=1}^N C_1^{\perp}$,
which implies $\HamW(L \setminus\{ \zrv \})/\Noa \le 1/N$,
where $\Noa$ and $N$ are the length of $L$ and that of the outer code,
respectively.
It was demonstrated that the underlying metric structure, $\HamD_{B}$,
plays a role in evaluating $\HamW(L\setminus B)$.

After completing the revision for the second submission,
the author learned that attainable asymptotic relative
minimum distance of concatenated quantum codes, where
the outer codes are CSS-type AG codes,
are also discussed in \cite{fujita06}.
However, the AG codes used in \cite{fujita06} 
are the non-constructible dual-containing codes specified in
\cite{rmatsumoto02}, and hence, the resulting codes are not constructible
%(Remark to Proposition~\ref{prop:enl}).
(cf.\ footnote 5). % of the present paper).
In \cite{fujita06}, 
symplectic codes from the table of \cite{crss98} are used as inner codes.
%concatenation of 
The best lower bound in \cite[Figure~2]{fujita06}, as ours,
depends on the parameters, $[[n,k,d]]$, of the inner code.
Unfortunately, 
these inner codes are not specified explicitly in \cite{fujita06}. However,
the plotted lines in \cite[Figure~2]{fujita06} suggest that
there seems to be only one choice of $[[n,k,d]]$ that gives
a line (lower bound) exceeding those given in the present work.
Namely,  in \cite[Figure~2]{fujita06}, 
one can find a lower bound, which is higher than
ours in the interval $0.071 \le \delta \le 0.10$, 
and which seems based on a non-CSS-type inner code.
The present author checked that this bound
can be attained by polynomially constructible codes replacing
the non-constructible outer codes in \cite{fujita06}
with the constructible codes used in the present work.

The issue of finding 
a polynomial construction of a tower of codes $D^{\perp} \le D \le D'$ with the optimal parameter $\gtNhat=\gtN$,
which was addressed in footnote 5 (Remark to Proposition~\ref{prop:enl}),
would be interesting.
This is because the enlarged CSS codes in Proposition~\ref{prop:enl} 
with $\gtNhat=\gtN$ outperform
the corresponding CSS codes,
and hence, improve on many of the best constructive bounds presented or mentioned
in this work.
This issue would be treated elsewhere.

The editor drew the author's attention to \cite[Section~7.3]{KWerner03},
where concatenation of a general quantum codes %(as inner codes)
and a `random graph code' %(as outer codes) 
was used in a Shannon-theoretic argument.
%cf.\ an analogous argument in \cite[Section~6]{hamada03f}.
However, complexity issues
%, which is the concern of the present work,  
were discarded in \cite{KWerner03}. %as usual in earlier works.

The title of the paper, largely suggested by the editor,
would be more suitable if 
the polynomial-time construction of efficiently decodable concatenated codes in \cite[Section~VI]{hamada06itw}
(where the restriction $k_1^{(i)} = k_2^{(i)}$ 
%on the inner code parameters can be dropped) were included.
on the inner codes can be dropped) had been included.
The codes achieve the same rate
$1-H(W_1)-H(W_2)$ %for %the channel pair $(W_1,W_2)$. 
as the codes in Theorem~\ref{th:performance} (Section~\ref{ss:goal}).

\appendices

%\appendix

\section{Proofs for Enlarged CSS Codes \label{app:enl}}

\subsection{Fixed-Point-Free Matrix \label{ss:fpfm}}

In this subsection, we show the existence of a needed fixed-point-free matrix.
In fact, it is a companion matrix defined in (\ref{eq:cm}).
Note that a fixed-point-free matrix
is a paraphrase of a matrix having no eigenvalue in $\myF$.

\begin{lemma}\label{lem:steaneAexist}
Let $\stA$ be (the transpose of) the companion matrix
of a polynomial $a(x)$ of degree 
$m \ge 2$ over $\myF$ that has no root in $\myF$.
Then, $\stA$ has no eigenvalue in $\myF$.
\end{lemma}

{\em Proof.}\/
The characteristic polynomial of $M$ is $a(x)$ itself
as can be checked by a direct calculation. 
Hence, $M$ has no eigenvalue in $\myF$.
\enproof

The next trivial fact shows that choosing such a polynomial $a(x)$
is a task of constant complexity in code-length.

\begin{lemma}\label{lem:steaneAexist2}
Suppose a polynomial $b_k(x)=x^k -a_{k-1} x^{k-1} - \cdots - a_1 x -a_0$
over $\myF$
has no root in $\myF$.
Then, for any integer $m \ge k$ with $m \equiv k$ $(\bmod \, q-1)$,
$b_m(x)=x^m -a_{k-1} x^{k-1} - \cdots - a_1 x -a_0$
has no root in $\myF$. 
\end{lemma}

\subsection{Proof of Lemma~\protect\ref{lem:steane}}

{\em Proof of Lemma~\ref{lem:steane} and its corollaries.}\/ 
We should only prove the bound on minimum distance
since the other part of the proof of \cite{steane99e} is valid for 
any prime power $\dmn$.

Denoting a generator matrix of $\genC'\mbox{}^{\perp}$ by $H'$,
we may assume $H'$ is a submatrix of the generator matrix $U$ of $\genC^{\perp}$.
%in (\ref{eq:Gpr}).
Then, since $\spn \cH \le \spn \cG$, we may assume
\[ %\begin{equation}
\cH' =  \left[ \begin{array}{c|c} 
H'   & 0 \\
0 & H'  \end{array}
\right]
\]
is a submatrix of the `stabilizer' matrix $\cH$, 
as shown in ~\cite{steane99e},
and hence is a submatrix of $\cG$ as well.

We consider $\HamW([u,v])$ for $x=(u|v)\in \spn\cG\setminus\spn\cH'$,
noting $\spn\cH'=\genC'\mbox{}^{\perp}\oplus \genC'\mbox{}^{\perp}$.
If no rows of $(V|\stA V)$ are involved in the generation of $(u|v)$,
then $\HamW([u,v]) \ge d$. Note, otherwise,
$u,v\in \genC'\setminus \genC'\mbox{}^{\perp}$ and $v\ne \lambda u$ for any $\lambda$.
Hence, we have the lemma. 

Corollary~\ref{coro:steane_CEL} immediately follows from the lemma.
We establish Corollary~\ref{coro:steane} by proving $d'' \ge 
\lceil \frac{\dmn+1}{\dmn} d' \rceil$. 
Namely, we show that 
for any pair of linearly independent vectors 
$u,v \in \genC' \setminus \genC'\mbox{}^{\perp}$,
we have $\HamW([u,v]) \ge \lceil \frac{\dmn+1}{\dmn} d' \rceil$. 
Write $u=(u_1,\dots,u_{\Noa})$, $v=(v_1,\dots,v_{\Noa})$,
and put $w=\HamW(u)$.
Without loss of generality,
we may assume $u_{w+1}=\cdots=u_{\Noa}=0$. 
Denoting the number of $i$ with $v_i = \lambda u_i$, $1\le i \le w$, 
by $l(\lambda)$ for $\lambda\in\myF$, we have an element $\lambda^*\in\myF$
with $l(\lambda^*) \ge w/\dmn$, the average of $l(\lambda)$. Then,
\[
d' \le \HamW(v -\lambda^* u) \le w - \frac{w}{\dmn} + 
\HamW\big( (v_{w+1},\dots,v_{\Noa}) \big).
\]  
Hence, we have
$\HamW([u,v]) = w+\HamW\big( (v_{w+1},\dots,v_{\Noa}) \big)
\ge d' +w/\dmn \ge d'(1+1/\dmn)$, and the corollary.
\enproof

\subsection{Proof of Proposition~\protect\ref{prop:enl}}

In our construction, we apply Lemma~\ref{lem:steane} % to this tower.
assuming the tower in (\ref{eq:tower}) is that in (\ref{eq:tower2}).
Note
$\dim C_1^{\perp} =(n-k)/2$, which follows from that $C_1/C_2^{\perp}$ is an
$[[n,k]]$ quotient code and $C_1=C_2$,
and hence,
\begin{gather*}
\Noa=nN, \quad \Koa = k \KoD + \frac{n-k}{2}N, \\
\Koa' = k \KoE+ \frac{n-k}{2}N
\end{gather*}
where
\[
\KoD=\dim_{\myFk} D, \quad \KoE=\dim_{\myFk} D'.
\]
Hence, the overall rate of the symplectic code is
\begin{equation}\label{eq:rateS}
\frac{\Koa+\Koa'-\Noa}{\Noa} = \frac{k}{n}\Big(\frac{\KoD+\KoE}{N}-1\Big).
\end{equation}

Put 
\[
\delta = \liminf_{\nu\to\infty} \frac{\HamW(\embone(D) \setminus B)}{\Noa},
\quad
\delta' = \liminf_{\nu\to\infty} \frac{\HamW(\embone(D') \setminus B)}{\Noa}.
\]
Then, 
the analysis in Section~\ref{ss:bound} that 
%by an analysis very similar to that in the previous section that 
leads to (\ref{eq:rmd1}) and (\ref{eq:rmd2}), 
which actually lower-bounds
the minimum distance of the concatenation of $C_j/C_{\bar{j}}^{\perp}$ and
$D_j/\{ \zrv \}=D_j$,
%$(C_1,C_2)$ and $(L_j, \myFkpower{N})$, not good since two cases involved
%$(C_1,C_2)$ and $(L_1, \myFkpower{N})$ 
% $(C_1,C_2)$ and $(L_2, \myFkpower{N})$ /
gives
\[
\delta \ge \frac{d}{n}(1 - \gtNhat -R ) \defeq \Delta, \quad
\delta' \ge \frac{d}{n}(1 - \gtNhat -R') \defeq \Delta'
\]
where $R,R'$ are the limits appearing in the condition (iii).

Putting 
\begin{equation}\label{eq:sys_eq}
R''=R+R'-1 \ \mbox{ and } \ \Delta=\Delta'(\dmn+1)/\dmn,
\end{equation}
we have
\[
\min\{ \delta, \delta'(\dmn+1)/\dmn \}  \ge 
\frac{(\dmn +1)d}{(2\dmn +1)n}( 1-2\gtNhat -R'').
\]
Then, noting (\ref{eq:rateS}) and
\[
\liminf_{\nu\to\infty} \frac{\KoD}{N} \ge R, \quad
\liminf_{\nu\to\infty} \frac{\KoE}{N} \ge R',
\]
which imply
\[
\liminf_{\nu\to\infty} \frac{\KoD + \KoE}{N} -1 \ge R'', \quad
\]
we see the overall rate of the symplectic code satisfies  
\[
\liminf_{\nu\to\infty} \frac{\Koa+\Koa'-\Noa}{\Noa} \ge \frac{k}{n} R'' = \Roa.
\]

Thus, the constructed $[[\Noa,\Koa'',\doa]]$ symplectic codes satisfy
\begin{equation}\label{eq:rateS2}
\liminf_{\nu\to\infty} \frac{\Koa''}{\Noa} \ge \Roa
\end{equation}
and
\begin{equation}\label{eq:boundSgen}
\liminf_{\nu\to\infty} \frac{\doa}{\Noa} \ge 
%l^{\rm enl}_t \defeq  
\frac{(\dmn+1)d}{2\dmn+1}\Big( \frac{1 -2\gtNhat}{n} -\frac{1}{k} \Roa \Big)
\end{equation}
by Corollary~\ref{coro:steane}.
Note (\ref{eq:boundSgen}) can be attained 
for any rate 
\begin{equation}\label{eq:Rgehalf}
\Roa \ge \frac{k}{2(\dmn+1)n}(1-2\gtNhat),
\end{equation}
which is a rewriting of $R \ge 1/2$.
(Given $\Roa$, put $R'' = n\Roa/k$ and
let $(R,R')$ be the solution
of (\ref{eq:sys_eq}); see also the remark to the proposition.)

\subsection{Proof of Lemma~\protect\ref{lem:dis2} \label{app:dis2}}

%{\small 
We prove this lemma by presenting a procedure for producing
generator matrices $G_n$ of the $[n,n-1]$ code
$C_1$ of properties (A') and (B') for $n=3,4,\dots$
recursively. The produced matrices $G_n$ will have 
the parity check vector $b$ in the first row.
Note $\myF$ has the subfield $\myFnoarg_4$
since $\dmn=2^{2m}$ for some $m\in\SNN$ by assumption.
Let $\zeta$ be a primitive element of this subfield.
The procedure starts with the following generator matrix $G_3$, 
which fulfills (A') and (B'), where $C_1=\spn G_3$ and $b$ equal 
to the first row of $G_3$: 
\[
G_3=
\left[ \begin{array}{cc|c}
\zeta & \zeta^2 & 1\\\hline
\zeta^2 & \zeta & 0
\end{array} \right].
\]

{\em Step 1 for $n=3$}.\/ Deleting the last column of $G_3$,
pasting $(0,0)$ at the bottom,
and pasting an appropriate $3 \times 2$ matrix on the right,
we have
\[
\tlG_4=
\left[ \begin{array}{cc|cc}
\zeta & \zeta^2 & \zeta & \zeta^2 \\\hline
\zeta^2 & \zeta & 0 & 0 \\
0 &  0          & \zeta^2 & \zeta
\end{array} \right],
\]
which has the desired properties (A') and (B') for $n=4$.

{\em Step 2 for $n=3$}\/. 
The matrix $\tlG_4$ can be changed,
by adding a scalar multiple of the first row to the last,
into 
%the following form
\[
G_4 =
\left[ \begin{array}{ccc|c}
\zeta & \zeta^2    & \zeta & \zeta^2 \\\hline
\zeta^2 & \zeta    & 0  & 0 \\
1 &  \zeta &  \zeta     & 0
\end{array} \right].
\]
(The change was made
so that the entries %of the matrix 
in the rightmost column vanishes except the 
uppermost entry.)
Obviously, this generator matrix also has the desired properties.
% for $n=4$.

For $n=4,5,\dots$, as well,
we can produce $\tlG_{n+1}$ and then
$G_{n+1}$ of the desired properties from $G_n$ 
repeating Steps 1 and 2, 
which generalizes for an arbitrary number $n \ge 3$.
The generalization is obvious except the choice
of the $n \times 2$ matrix in Step 1. This matrix should be the transpose of
\[
\left[ \begin{array}{ccccc}
\lambda \zeta   & 0 & \cdots & 0 & \zeta^2 \\
\lambda \zeta^2 & 0 & \cdots & 0 & \zeta   
\end{array} \right]
\]
where $\lambda$ is the $(1,n)$-entry of $G_{n}$,
which is needed to make the first row of $G_{n+1}$ self-orthogonal. 
Thus, we have the desired generator matrices $G_n$ of $[n,n-1]$
codes $C_1$ for $n\ge 3$.
%}

\end{document}